\documentclass[showpacs, prb,twocolumn,preprintnumbers , superscriptaddress, aps]{revtex4-2}
 
\usepackage{color}
\usepackage{amsmath,amssymb}
\usepackage{pifont}
\usepackage{amssymb}  
\usepackage{bbold}
\usepackage{float}
\usepackage[countmax]{subfloat}

\usepackage{subfloat}
\usepackage{amsmath,amssymb}
\usepackage{mathtools}
\usepackage{physics}
\usepackage{breqn} 

\usepackage[caption=false]{subfig}
\usepackage{tikz}
\usepackage{makecell}
\usepackage{subfig}
\usepackage{pifont}   
\usepackage{graphicx} 
\usepackage{dcolumn}  
\usepackage{bm}       
\usepackage{multirow} 
\usepackage{placeins}
\usepackage[colorlinks]{hyperref}

\usepackage{mathtools}

\usepackage{breqn}
\usepackage{mathrsfs}

\usepackage{multirow}

\let\oldbibitem\bibitem
\renewcommand{\bibitem}{\vspace{4pt}\oldbibitem}

\begin{document}

\title{Flat and Topological Floquet Minibands from Patterned Light in Untwisted Bilayer Graphene}

	\author{Muhammad Faisal}
	\affiliation{Department of Physics$,$ Quaid-i-Azam University$,$ Islamabad$,$ 45320$,$ Pakistan}
    \author{Michael Vogl}
		\affiliation{Physics Department$,$
		King Fahd University
		of Petroleum $\&$ Minerals$,$
		Dhahran 31261$,$ Saudi Arabia}
      \affiliation{Interdisciplinary Research Center (IRC) for Advanced Quantum Computing$,$ KFUPM$,$ Dhahran$,$ Saudi Arabia}

\begin{abstract}
 Two-dimensional superlattices in van der Waals materials host flat bands and nontrivial topology, most famously at the so-called magic angles of twisted bilayer graphene, where flat bands give rise to correlated and topological phases. Yet these superlattices are usually created by twisting the layers or applying strain, and once a sample is fabricated, their period is fixed and extremely difficult to tune. Here we propose an alternative route: imprint the superlattice optically, using patterned electromagnetic fields rather than a physical twist or strain. We show that patterned in-plane circularly polarized light and a combined drive consisting of a patterned out-of-plane longitudinal field and a uniform circularly polarized field produce isolated bands in both AA- and AB-stacked bilayer graphene. In AB stacking, the central bands additionally become nearly flat, capturing key features of a driven moir\'e superlattice. In this approach, the superlattice period is set by the illumination and is straightforward to tune, and circularly polarized light breaks time-reversal symmetry. Computing the valley Chern numbers of the central bands, we find a rich topological structure with several phase transitions in both stackings. Our results establish light-induced superlattices as a flexible and tunable platform for engineering flat bands and topological phases in bilayer graphene.
\end{abstract}

\maketitle

\section{Introduction}

Superlattices are an effective means of controlling the electronic properties of quantum materials~\cite{esaki1970superlattice,geim2013van,wallbank2015moire}. They introduce a periodic potential with a length scale much larger than the underlying crystal lattice, folding the Brillouin zone into a mini-Brillouin zone and reconstructing the electronic spectrum into minibands~\cite{forsythe2018band,li2021anisotropic}. This reconstruction combines states that were originally separated in momentum space and produces minibands, some of which become relatively isolated and exhibit substantially reduced bandwidth. It also modifies the hybridization between these states, leading to the opening and closing of energy gaps at the mini-Brillouin-zone boundaries~\cite{wallbank2015moire,park2008anisotropic}. The reconstructed miniband structure also provides a natural platform for engineering topological phases using perturbations, such as periodic driving.~\cite{oka2009,lindner2011floquet,rudner2020band}. The most prominent realization of this idea is in twisted bilayer graphene, where a small relative rotation between two layers produces a moir\'e superlattice whose period is set by the twist angle~\cite{bistritzer2011moire}. Near the so-called magic angles, the resulting bands become  nearly flat, giving rise to correlated insulating~\cite{Cao2018Correlated}, superconducting~\cite{cao2018unconventional} and topological~\cite{sharpe2019emergent} phases. Similar behavior is found in other twisted materials, such as twisted transition-metal dichalcogenides~\cite{wang2020correlated}. However, there is a growing interest in realizing similar properties without twisting, as the precise control of the twist angle is still experimentally challenging~\cite{uri2020mapping}. The superlattice period is highly sensitive to the twist angle, so small fabrication errors can substantially alter the resulting band structure. In addition, the geometry of the superlattice is locked upon assembly, leaving little room for post-fabrication tuning~\cite{lau2022reproducibility}. One way to avoid the twist is to use materials whose superlattice comes from a lattice mismatch, such as graphene aligned with hexagonal boron nitride. The period is then fixed by the ratio of the two lattice constants, which are properties of the materials rather than something set during assembly, so the synthesis is more reliable. Such untwisted platforms have been shown to host much of the same physics, including band flattening and nontrivial topology under irradiation~\cite{Alabdulal_2025}. Another approach is to create a superlattice by introducing a periodic deformation of the crystal lattice. A deformation with a chosen symmetry enlarges the unit cell and folds the Brillouin zone, resulting in band hybridization and gap opening at the new zone boundaries. When time-reversal symmetry is broken, the reconstructed bands can acquire nonzero Chern numbers~\cite{aljishi2026bandstructuretopologyperiodically,hammadi2026magnon}. This approach is nevertheless limited in flexibility, since the material's mechanical stability constrains the deformation pattern. Once the strain is applied, the superlattice geometry is essentially fixed.

These limitations motivate the exploration of more flexible and externally tunable approaches to superlattice engineering. One possibility is to project patterns directly onto a material, for example using light, to mimic the superlattice behavior of materials without twist. This idea is related to Floquet engineering, where a system is driven by a time-periodic field, allowing its electronic band structure and topology to be modified without changing the material itself~\cite{oka2009,lindner2011floquet,rudner2020band}. In moiré superlattices, periodic driving has been shown to flatten minibands, induce topological phase transitions, and alter the associated Chern numbers~\cite{Topp2019,Li2020}. Patterned light offers considerably greater flexibility because the superlattice period and symmetry are determined by the spatial profile of the applied field rather than by the material itself. It is therefore interesting to explore whether such patterned illumination can be used to mimic the superlattice behavior of materials without twist. This approach opens a new direction for engineering electronic and topological properties, and it is the route we pursue here. In particular, we compare different forms of light irradiation to mimic the band structure and topological phases of driven and undriven moir\'e superlattices without twist in bilayer graphene. In this work, we compare three driving schemes. The first is patterned circularly polarized light, the second is patterned longitudinal light, and the third is patterned longitudinal light combined with a spatially uniform circularly polarized drive. These driving schemes are expected to modify the electronic structure differently, as circularly polarized and longitudinal light couple to bilayer graphene through different physical mechanisms. By comparing them, we can identify their roles in shaping the band structure and the resulting topological phases.

The remainder of this paper is organized as follows. In Sec.~\ref{setups}, we introduce the proposed experimental setups and the theoretical framework. We derive the vector potentials for the patterned in-plane and longitudinal driving configurations, present the $k\cdot p$ Hamiltonians for AA- and AB-stacked bilayer graphene, describe their coupling to light, and outline the Floquet-Bloch formalism together with the convergence procedure used in the numerical calculations. In Sec.~\ref{results}, we first present the quasienergy band structures at increasing driving strengths for fixed superlattice period and frequency, highlighting the isolation and flattening of the central bands. We then analyze the corresponding effective Floquet Hamiltonians obtained from the Van Vleck expansion to clarify the origin of the light-induced band reconstruction. Finally, we investigate the topological properties of the isolated bands by constructing valley Chern-number phase diagrams for the relevant driving configurations and identifying the associated topological phase transitions. Additional results on the frequency and superlattice-period dependence are presented in the Appendix.

\section{Suggested Setup and Physical model}
\label{setups}

In this section, we describe our setups for bilayer graphene subjected to patterned light. In particular, we consider three cases: patterned circularly polarized light, patterned longitudinal light, and uniform circularly polarized light plus longitudinal light. We derive the vector potentials and Hamiltonians and discuss the numerical implementation.

\subsection{Envisioned experimental setups}

The main aim of this work is to investigate bilayer graphene illuminated by spatially patterned light and understand its effects on band structure and band topology. Our work is mainly driven by the question of whether it is possible to obtain physics similar to irradiated twisted bilayer graphene or other moir\'e materials, albeit without a twist\cite{PhysRevResearch.1.023031, PhysRevResearch.2.033494}.

Here, the initial objective is first to understand how spatially patterned optical fields influence both the intralayer and interlayer couplings in bilayer graphene. Our approach is guided by the idea that closely emulating certain features of the couplings from twisted bilayer graphene might lead to similarly interesting physics. For instance, in TBG the moir\'{e} superlattice gives rise to a spatial modulation of the interlayer coupling\cite{Bistritzer2011,Cao2018Correlated,LopesDosSantos2007}. Rather than introducing a physical twist between the graphene layers, we propose to employ patterned light. In our work, we investigate whether similar spatial modulation of both couplings can be achieved optically and examine its influence on the electronic band structure. We limit our study to a hexagonal illumination pattern to best mimic the symmetry of the moir\'{e} superlattice. Therefore, in this work, we generate all patterned light fields using optical setups with hexagonal masks, creating a periodic optical field with hexagonal symmetry.

Throughout this work, we keep the squared ratio between the aperture diameter and the hole
spacing fixed at $(d/L)^2$ = 0.36, with the smallest pattern given by a diameter d = 60 nm and a spacing L = 100 nm, because this squared ratio determines the fraction of open area in the mask and therefore the transmitted light intensity. Keeping it fixed means the transmitted intensity stays approximately the same (up to diffraction and aperture-resonance effects) as the pattern is rescaled, so only the superlattice period changes, making the result easier to interpret. We choose the smallest diameter to be d = 60 nm because it is comfortably larger than the resolution limit of electron-beam lithography, which is on the order of 10 nm. Hence, the pattern is easier to fabricate, and nanoimprint lithography can reproduce the same features over large areas\cite{li2000fabrication,maximov2002fabrication}.
This provides a practical experimental knob for investigating how the electronic band structure evolves as a function of the spatial periodicity of the patterned light.
In this work, we will pit three different forms of patterned light against each other and compare their effects. In this section, we focus on the setup that generated each form of light; we develop a detailed mathematical description in the subsequent section. This approach separates parts of the work that are mostly interesting to theorists from those more interesting to a general audience.

\begin{figure}
    \centering
\includegraphics[width=\columnwidth]{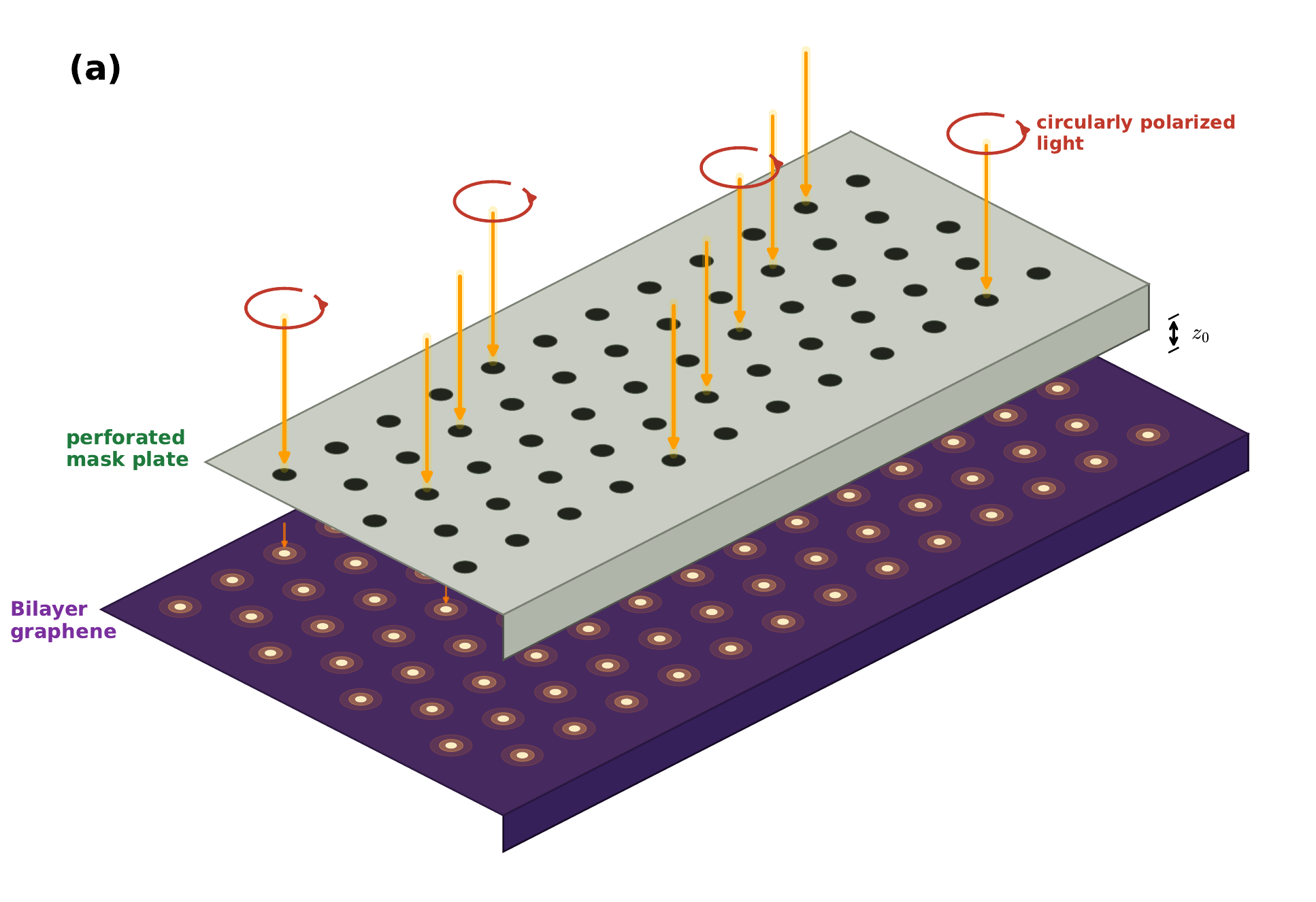}
    \caption{ Cartoon of bilayer graphene illuminated by spatially patterned circularly polarized light transmitted through a hexagonal mask.} 
   \label{case_1}
\end{figure}

We first consider a configuration in which circularly polarized light is transmitted through the hexagonal mask and illuminates the bilayer graphene as shown in Fig.~\ref{case_1}.
In this configuration, circularly polarized light couples to in-plane momenta of electrons via the so-called minimal coupling procedure~\cite{chebrolu2019flat,marino2015interaction}, while the hexagonal mask introduces a spatially periodic modulation of the incident optical field.\\

\begin{figure}
    \centering
    \includegraphics[width=\columnwidth]{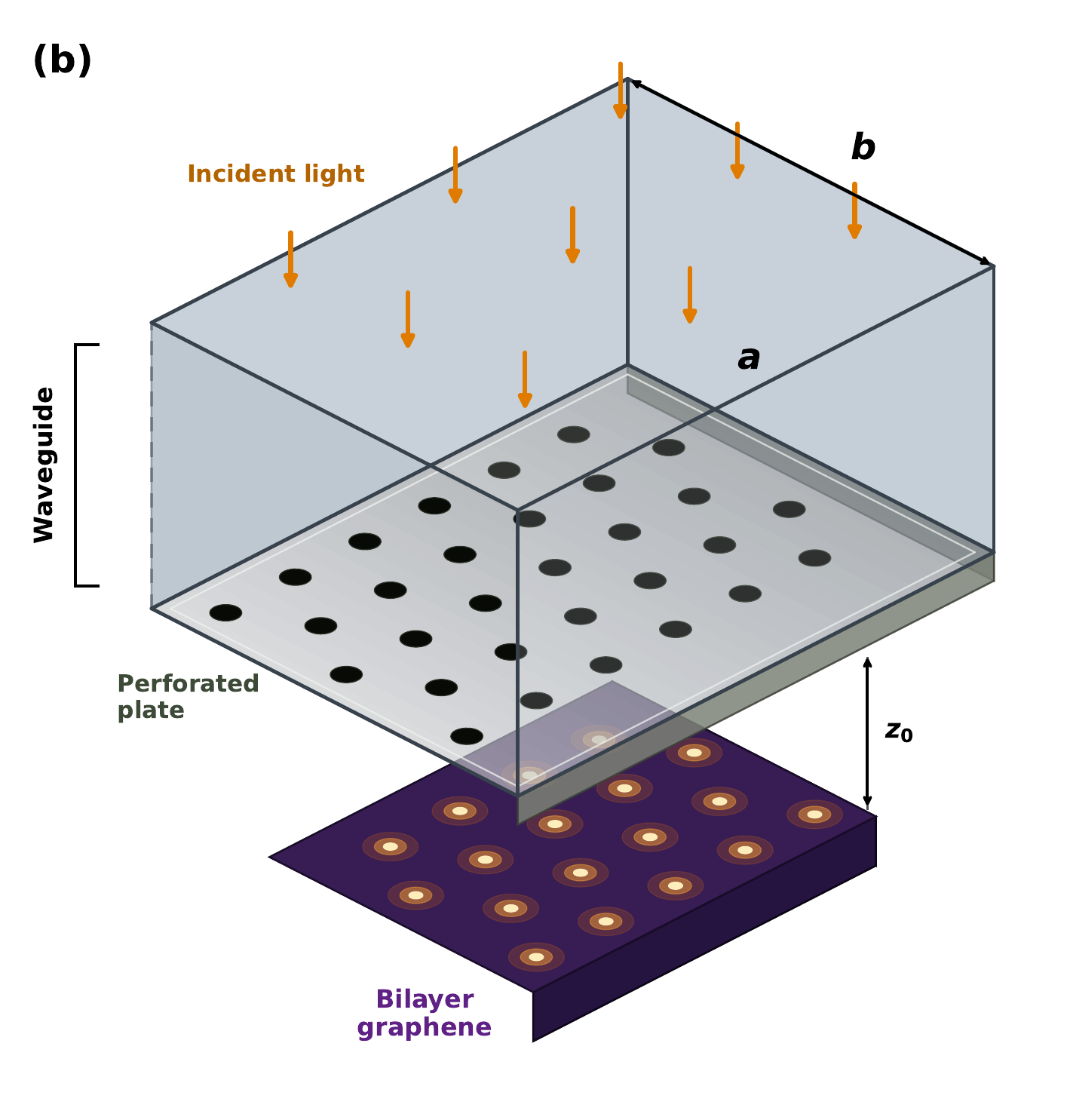}
    \caption{Cartoon for illuminating bilayer graphene with a spatially patterned longitudinal light field generated by a vertical waveguide with a hexagonal array of apertures at the exit of the waveguide.}  
   \label{case_2}
\end{figure}

Second, we consider a setup designed to modify interlayer couplings - most closely mimicking twisted moir\'e materials. Since circularly polarized light couples through minimal coupling and only affects the intralayer hopping, a slightly unconventional approach is necessary. In principle, one could try to use normal light at a glancing angle of incidence. Tilting the incident light so that it propagates nearly parallel to the bilayer graphene 
produces an out-of-plane vector-potential component, $A_z = A_0 \sin\theta$, where $\theta$
is the angle of incidence measured from the sample normal. Such a component can, in principle, couple to the interlayer hopping. However, the power
delivered to the sample is proportional to the normal component of the Poynting vector,
$S_\perp = S_0 \cos\theta$, which vanishes in the grazing-incidence limit
$(\theta \rightarrow 90^\circ)$. Consequently, although $A_z$ is enhanced, the energy
reaching the sample becomes very small~\cite{jackson1999}. A more natural idea is to use light whose field points directly along the stacking direction: longitudinal light. In free space, this is not possible because Maxwell's equations allow only transverse electromagnetic waves and forbid a longitudinal field component. To obtain such a longitudinal component, one has to use a rectangular waveguide, which is exactly what our aperture setup provides. The light
transmitted through the sub-wavelength holes has a field component along the stacking
direction that is absent for ordinary light in vacuum, and this is what allows us to
drive the interlayer coupling. In particular, we introduce longitudinal light, which affects the interlayer hoppings directly, as discussed in ~\cite{ohta2012evidence,wang2016stacking}.  A patterned version of such a field can be generated by a rectangular waveguide with a hexagonal mask at the exit of the waveguide, as shown in Fig.~\ref{case_2}.

We choose the waveguide cross-section to be $a=12~\mu\mathrm{m}$ and $b=13~\mu\mathrm{m}$, so that the longitudinal component $A_z$ varies slowly near the center of the waveguide cross-section. The longitudinal component $A_z$ remains within $5\%$ of its maximum over an area covering approximately
$24 \times 26$ superlattice periods for $L = 100~\mathrm{nm}$, and $12 \times 13$ periods for the
largest superlattice period considered, $L = 200nm~\mathrm{nm}$. The transverse mode profile sets this uniform region and can
be enlarged simply by increasing the cross section, which extends the
uniform-field area to cover more superlattice periods. Increasing the size only
lowers the cutoff frequency $\hbar\omega_c = \pi\hbar c\sqrt{1/a^2 + 1/b^2}$,
which poses no problem here since we always operate at higher frequencies. This
gives ample flexibility to enlarge the uniform-field region for experiment. Thus, the longitudinal component is approximately uniform over this region. The complete waveguide mode also contains transverse components away from the center; in the electronic model, we retain only $A_z$ and treat the longitudinal drive as an idealized effective protocol.  Our Hamiltonian is periodic in both space and time, as we will see as we progress, and its spectrum spans a wide range of energies.


For Floquet engineering, the driving frequency must exceed the spectral width of the truncated Hamiltonian used in the numerical calculations. Although the $k \cdot p$ Hamiltonian is
formally unbounded, the Floquet Hamiltonian becomes finite after truncation to a finite number
of reciprocal-lattice shells. Throughout this work, we employ a maximum truncation of
$N_s = 5$ shells, for which the spectral width is approximately $0.91$~eV for AB stacking and
$0.77$~eV for AA stacking. Every frequency we use lies above the cutoff of this cross-section,
$\Omega_c = 0.07~\mathrm{eV}$, and the longitudinal mode is always available. The patterned apertures just below the waveguide produce a spatially periodic longitudinal light field, and the bilayer graphene is positioned within $z=10 \text{ nm}$ of the bottom plate to ensure patterned near-field illumination and prevent evanescent decay. The placement of the graphene sheets at 10 nm below the perforated plate in vacuum means that it is close enough that roughly
half to two-thirds of the pattern amplitude survives at the sample, with the details given
in the next section.\\ 

\begin{figure}[H]
    \centering
    \includegraphics[width=\columnwidth]{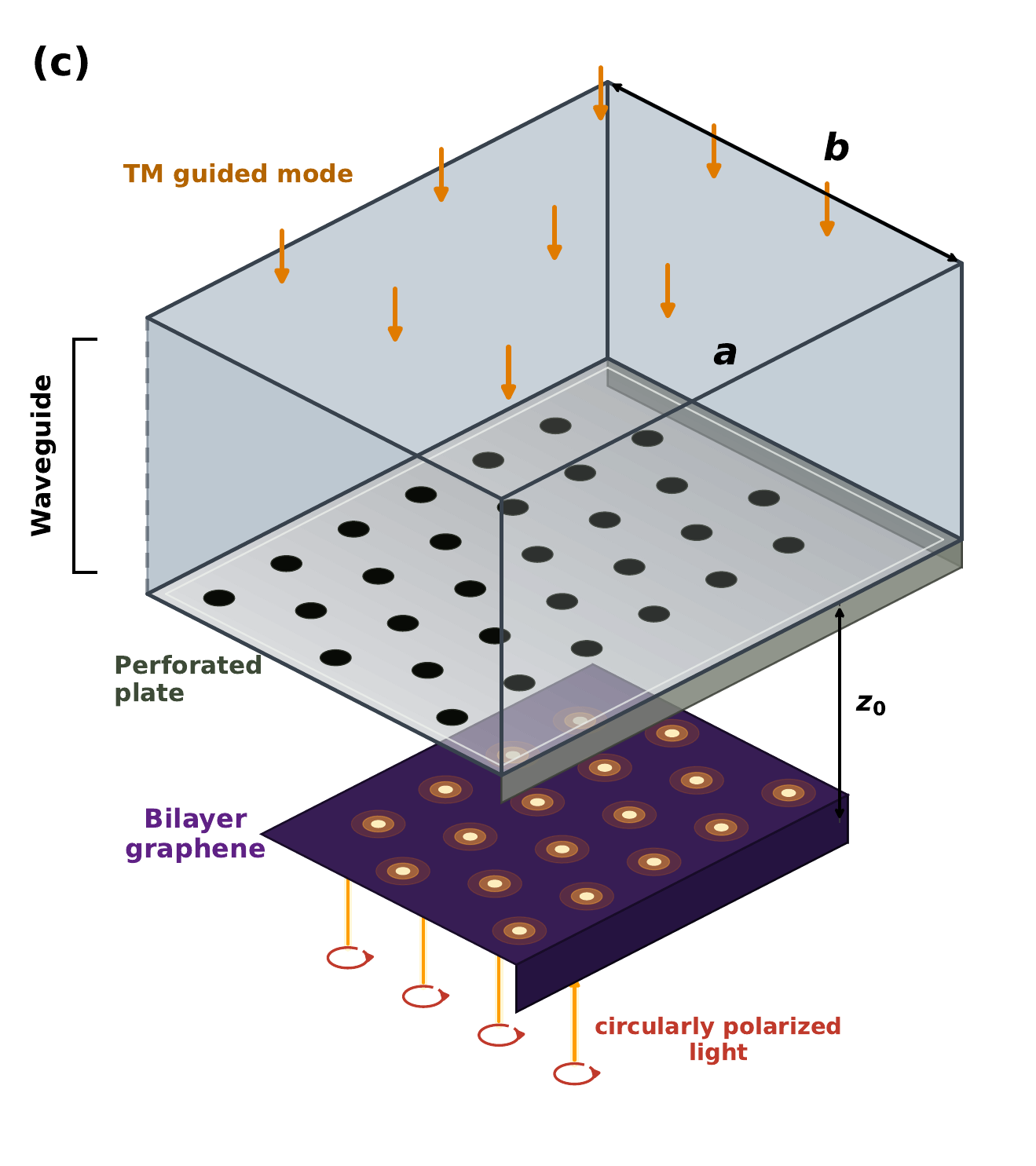}
    \caption{Cartoon of simultaneous illumination of bilayer graphene with uniform circularly polarized and  patterned longitudinal light generated by a rectangular waveguide as  described in Case~\ref{case_2}.} 
   \label{case_3}
\end{figure}

Finally, as we will see later, it is more interesting to study the simultaneous application of uniform circularly polarized light and patterned longitudinal light through the experimental setup shown in Fig.~\ref{case_3}, so that the intralayer and interlayer couplings can be modulated at the same time.
This allows us to control the intralayer and interlayer coupling channels independently and to investigate their combined effect on the electronic band structure of bilayer graphene, as well as the interplay of the two optical driving mechanisms. Most importantly, gaps opened by circularly polarized light permit investigations into band topology and allow us to mimic the behavior of irradiated moire materials most closely.

\subsection{Patterned light and its vector potential}
After discussing possible experimental setups for our study, we now turn to the underlying mathematical formulation.

First, we describe each optical field by its vector potential $\mathbf{A}(\mathbf{r},t)$, because this is the quantity that enters Hamiltonians most naturally. For a layered 2D material such as bilayer graphene, intralayer motion couples to $\mathbf{A}$ through the minimal substitution
$\hbar\mathbf{q}\to\hbar\mathbf{q}+e\mathbf{A}$, where $e>0$ is the magnitude of the electron charge, whereas the interlayer hopping couples through a Peierls phase. We therefore obtain $\mathbf{A}(\mathbf{r},t)$ for each type of illumination from Maxwell’s equations within a realistic but simplified model of the experimental geometry.
We first consider circularly polarized light without the mask. This light is transverse, so its vector potential lies in the plane of the bilayer and rotates in time according to
\begin{equation}
\mathbf{A}(t) = A_0\left(\cos\Omega t\,\hat{\mathbf{x}} \pm \sin\Omega t\,\hat{\mathbf{y}}\right).
\end{equation}
Here, $A_0$ is the amplitude and $\Omega$ is the drive frequency. The relative sign of the two terms sets whether the light is left- or right-circularly polarized. Here, we assume that the wavelength of the light is much larger than the graphene lattice constant, so the field has the same strength everywhere in the sample. Therefore, $\mathbf{A}$ does not change for different layers~\cite{karch2010dynamic,park2011visualizing}.\

We now determine the transmitted patterned vector potential generated by circularly polarized light passing through the hexagonal mask in Fig.~\ref{case_1}. To determine the transmitted field, we solve Maxwell’s equations in the region above and below the mask subject to the aperture boundary conditions.
We model the hexagonal mask as a perfect electric conductor at $z=0$ perforated by circular
apertures of radius $r$ arranged with hexagonal symmetry of period  $L$. The mask is illuminated from
above by circularly polarized light, and the bilayer graphene is at a distance $z=-z_{0}$.

The transmitted field just below the mask can be modeled, using the Kirchhoff condition~\cite{kirchhoff1883,karch2010dynamic}, by
multiplying the incident circularly polarized light by an aperture function
$T(\mathbf{r}\parallel)$, which equals unity inside the holes and zero on the metal.
\begin{equation}\label{CPL Potential}
\mathbf{A}(\mathbf{r}_\parallel, 0, t) = A_0\, T(\mathbf{r}_\parallel)
\big( \cos\Omega t\, \hat{x} + \sin\Omega t\, \hat{y} \big).
\end{equation}
Since the holes are arranged periodically, the aperture function is periodic in the plane and may be expanded in a Fourier series over the reciprocal lattice vectors of the mask\cite{garcia2010light},
\begin{equation}
T(\mathbf{r}_\parallel) = \sum_{\mathbf{G}} T_{\mathbf{G}}\,
e^{i\mathbf{G} \cdot \mathbf{r}_\parallel} ,
\label{Fourier Transform}
\end{equation}
where ${\mathbf{G}}$ is the set of reciprocal lattice vectors of the mask. The first shell of reciprocal lattice vectors consists of the six vectors $\mathbf G_i$ with length
$|\mathbf G_i|=4\pi/(\sqrt{3}L)$. The coefficients $\mathbf{T}_{\mathbf{G}}$ are the amplitudes of the Fourier components of the transmitted field, fixed by the boundary conditions at the apertures. They determine how strongly each spatial harmonic of the aperture array is imprinted on the
light, and it is these components that determine details of the periodic superlattice. We can determine these Fourier coefficients $T_{\mathbf{G}}$ from the orthogonality of the
plane waves over one unit cell of the mask,
\begin{equation}
\int_{\rm cell} e^{i(\mathbf{G} - \mathbf{G}') \cdot \mathbf{r}_\parallel}\, d^2 r
= A_{\rm uc}\, \delta_{\mathbf{G}, \mathbf{G}'} ,
\end{equation}
where $A_{\rm uc} = \sqrt{3} L^2/2$ is the area of the unit cell of a triangular (hexagonal) lattice with lattice constant L. Multiplying Eq.~\ref{Fourier Transform} by
$e^{-i\mathbf{G}’ \cdot \mathbf{r}\parallel}$, using the orthogonality condition,  and integrating over the cell gives
\begin{equation}
T_{\mathbf{G}} = \frac{1}{A_{\rm uc}} \int_{\rm cell} T(\mathbf{r}_\parallel)\,
e^{-i\mathbf{G} \cdot \mathbf{r}_\parallel}\, d^2 r
= \frac{1}{A_{\rm uc}} \int_{\rm hole} e^{-i\mathbf{G} \cdot \mathbf{r}_\parallel}\, d^2 r ,
\end{equation}
where the last step follows because the aperture function vanishes on the metal and equals
unity inside the holes, so that the integral over the unit cell reduces to an integral over
a single aperture. For circular apertures of radius $r$ the integral can be evaluated in polar coordinates, and
gives
\begin{equation}
T_0 = \frac{\pi r^2}{A_{\rm uc}} ,
\qquad
T_{\mathbf{G}} = T_0\, \frac{2 J_1(|\mathbf{G}| r)}{|\mathbf{G}| r} ,
\end{equation}
where $J_1$ is the Bessel function of the first kind. The uniform coefficient $T_0$ is simply
the fraction of the surface that is open, since for $\mathbf{G} = 0$ the exponential reduces
to unity and the integral measures the area of a single hole relative to the area of the unit
cell. Since we keep the ratio $(d/L)^2$ fixed at $0.36$, the argument of the Bessel function is
independent of the pattern size. One has $|\mathbf{G}| r = 2.18$ for the first shell, so that
$T_0 = 0.33$ and the contrast $\alpha \equiv T_{\mathbf{G}}/T_0 = 0.51$ take the same values
for every period $L$ considered. Substituting the Fourier expansion of the aperture function into Eq.~\ref{CPL Potential}, the
vector potential
\begin{equation}
\begin{split}
\mathbf{A}(\mathbf{r}_\parallel, z, t) = {} & A_0 \Big[ T_0
+ \sum_{\mathbf{G} \neq 0} T_{\mathbf{G}}\, e^{i\mathbf{G} \cdot \mathbf{r}_\parallel}\,
 \Big] \\
& \times \big( \cos\Omega t\, \hat{x} + \sin\Omega t\, \hat{y} \big)
\end{split}
\end{equation}
This is the boundary
condition from which the field below the plate is determined.
The transmitted field necessarily contains an out-of-plane component $A_z$, as discussed below,
This follows from working in the Coulomb gauge since a patterned in-plane field alone is not divergence-free. We note that the out-of-plane component $A_z$ couples only to the interlayer
hopping, since the hopping between the layers is along $\hat{z}$ and only $A_z$
points that way. Through the Peierls substitution, the hopping acquires a phase
$\phi = (e/\hbar)\int \mathbf{A}\cdot d\boldsymbol{\ell} = (e/\hbar) A_z, d$,
where $d$ is the interlayer spacing (in our case $d=0.34$~nm)~\cite{guinea2019}
and $\phi$ is dimensionless. The Coulomb gauge condition fixes its size
$\nabla\cdot\mathbf{A} = 0$, which gives
$\partial_z A_z = -(\partial_x A_x + \partial_y A_y)$, so that $A_z$ is sourced
only by the in-plane variation of the field. The uniform component of the
in-plane field is constant in the plane and therefore generates no $A_z$; only
the patterned component, of relative weight $M(L)$, contributes. Since the
pattern is built from first-shell plane waves, taking its gradient contributes a
factor $|\mathbf{G}|$, while $A_z$ decays along $z$ as
$\partial_z A_z \sim \kappa{\mathbf{G}} A_z$, and balancing the two gives
$A_z \sim (|\mathbf{G}|/\kappa_{\mathbf{G}}) M(L) A_0 T_0$. In the subwavelength
regime $\kappa_{\mathbf{G}} \approx |\mathbf{G}|$, so $A_z \sim M(L), A_0 T_0$,
and with $a_0 = (e/\hbar) A_0 T_0$, the Peierls phase becomes
$\phi \sim \phi_A = M(L), a_0, d$, where $\phi_A$ sets the amplitude of any
contribution. For the strongest drives used in this work, $a_0 = 0.30$ for AA and
$a_0 = 0.15$ for AB, the resulting Peierls phase reaches at most
$\phi_A \approx 0.07$ and $0.04$, respectively, so the neglected interlayer phase
remains small in all cases. We have verified numerically that including this
phase leaves the isolated flat bands and their gaps unchanged, shifting the
central bands by less than $10^{-3}$~meV, so we neglect the component $A_z$ in our
discussion and retain only the in-plane minimal coupling.\

In the region between the plate and the graphene, we work in the Coulomb gauge, defined by
$\nabla \cdot \mathbf{A} = 0$. This region is source-free vacuum, so Gauss’s law reads
$\nabla \cdot \mathbf{E} = 0$, and substituting $\mathbf{E} = -\nabla\varphi - \partial_t \mathbf{A}$
together with the gauge condition gives $\nabla^2 \varphi = 0$. In the absence of an externally applied scalar potential, we choose the solution $\varphi=0$. This additional choice is consistent with the Coulomb gauge and the source-free Maxwell equations; it does not follow from the Coulomb-gauge condition alone. The electric field is then determined by the vector potential alone,
$\mathbf{E} = -\partial_t \mathbf{A}$, and Gauss’s law is satisfied identically since
$\nabla \cdot \mathbf{E} = -\partial_t(\nabla \cdot \mathbf{A}) = 0$. Maxwell’s equations
reduce to the Helmholtz equation
\begin{equation}
\big( \nabla^2 + k^2 \big) A_i = 0 , \qquad k = \Omega/c ,
\end{equation}
which each Cartesian component of $\mathbf{A}$ satisfies independently.
We now solve this equation in the region below the mask, subject to the boundary condition at $z = 0$ derived above. Since the boundary condition at $z = 0$ is periodic in the plane, the field below the mask must carry the same periodicity.
We stress that while the field emerging from a single aperture does spread sideways, a periodic array of apertures
cannot produce a new spatial period. The aperture coefficients $T_{\mathbf{G}}$ fix how the field is
distributed among the reciprocal-lattice harmonics $\mathbf{G}$ at the plate, and away from the
plate each harmonic falls off at its own rate. As we see below, that rate grows with $|\mathbf{G}|$,
so the higher harmonics die out first and the pattern grows smoother with distance while its period
stays $L$. The only departure from exact periodicity comes from the finite size of the array, and it
is confined to an edge region a few decay lengths wide, which we estimate below to be about
$0.14L$. This is negligible for a sample sitting well inside an array that spans many periods.
The field below the plate therefore keeps the same in-plane Fourier components as at the plate, and
differs only in how each one varies along $z$. Assigning to each harmonic its own profile
$Z_{\mathbf{G}}(z)$, normalised so that $Z_{\mathbf{G}}(0) = 1$, Eq.~(7) becomes

\begin{equation}
\begin{split}
\mathbf{A}(\mathbf{r}_\parallel, z, t) = {} & A_0 \Big[ T_0Z_{\mathbf{0}}(z)\
+ \sum_{\mathbf{G} \neq 0} T_{\mathbf{G}}\, e^{i\mathbf{G} \cdot \mathbf{r}_\parallel}\,Z_{\mathbf{G}}(z)\
 \Big] \\
& \times \big( \cos\Omega t\, \hat{x} + \sin\Omega t\, \hat{y} \big)
\end{split}
\end{equation}
where $Z_{\mathbf{G}}(z)$ is the profile of each Fourier component along the direction normal
to the mask, still to be determined, and $Z_{\mathbf{G}}(0) = 1$ so that the boundary condition
at the plate is recovered. Substituting this into the Helmholtz equation, so that each Fourier
component satisfies a one-dimensional equation in $z$,
\begin{equation}
\frac{d^2 Z_{\mathbf{G}}}{dz^2} - \kappa_{\mathbf{G}}^2\, Z_{\mathbf{G}} = 0 ,
\qquad
\kappa_{\mathbf{G}} = \sqrt{|\mathbf{G}|^2 - k^2} ,
\end{equation}
The solutions of this equation are $Z_{\mathbf{G}}\propto e^{\pm\kappa_{\mathbf{G}}z}$. Over the parameter range considered here, $\hbar\Omega=1$–$3$~eV and $L=100$–$180$~nm, every nonzero reciprocal-lattice vector retained in the calculation satisfies $|\mathbf{G}|>k$. The corresponding components are therefore evanescent, with $1/\kappa_{\mathbf{G}}$ setting their decay length. For the first reciprocal-lattice shell, $\kappa_{\mathbf{G}}/|\mathbf{G}|\geq0.91$, so we use the approximation $\kappa_{\mathbf{G}}\simeq|\mathbf{G}|$.
This is what
fixes the placement of the graphene in Fig.~\ref{case_1}, which must sit close enough to the
mask that the patterned components have not decayed away. For $\mathbf{G} = 0$, the situation is reversed. Here
$\kappa_0 = \sqrt{-k^2} = i k$ is imaginary, so the solution $e^{\pm i k z}$ oscillates rather
than decays, and this component propagates down to the sample. Since $kz_0\lesssim0.15$ over the small standoff considered here, this component acquires only a small propagation phase, which we neglect.
Collecting both cases, and keeping the solutions that do not grow away from the mask, the
transmitted field below the plate is
\begin{equation}
\begin{split}
\mathbf{A}(\mathbf{r}_\parallel, z, t) = {} & A_0 \Big[ T_0\, e^{-ik|z|}
+ \sum_{\mathbf{G} \neq 0} T_{\mathbf{G}}\, e^{i\mathbf{G} \cdot \mathbf{r}_\parallel}\,
e^{-\kappa_{\mathbf{G}}|z|} \Big] \\
& \times \big( \cos\Omega t\, \hat{x} + \sin\Omega t\, \hat{y} \big) \\[4pt]
\simeq {} & A_0 \Big[ T_0
+ \sum_{\mathbf{G} \neq 0} T_{\mathbf{G}}\, e^{i\mathbf{G} \cdot \mathbf{r}_\parallel}\,
e^{-\kappa_{\mathbf{G}}|z|} \Big] \\
& \times \big( \cos\Omega t\, \hat{x} + \sin\Omega t\, \hat{y} \big) ,
\end{split}
\end{equation}
where every patterned term is suppressed by its own evanescent
factor.
We approximate the Fourier sum by retaining the first reciprocal-lattice shell, which captures the leading hexagonal harmonic of the transmitted field. The higher spatial harmonics decay more rapidly away from the plate, and neglecting them gives the smooth optical-superlattice profile used in the numerical calculations. The six
first-shell vectors $\pm\mathbf{G}_{1,2,3}$ all have the same magnitude, and therefore the same
coefficient $T_{\mathbf{G}} = \alpha T_0$ and the same decay constant
$\kappa_{\mathbf{G}} \simeq G$, so that
\begin{equation}
\begin{split}
\mathbf{A}(\mathbf{r}_\parallel, z, t) = {} & A_0 T_0 \Big[ 1
+ \alpha\, e^{-G|z|} \sum_{n=1}^{3} \big( e^{i\mathbf{G}_n \cdot \mathbf{r}_\parallel}
+ e^{-i\mathbf{G}_n \cdot \mathbf{r}_\parallel} \big) \Big] \\
& \times \big( \cos\Omega t\, \hat{x} + \sin\Omega t\, \hat{y} \big) .
\end{split}
\end{equation}
Combining each pair of opposite vectors into a cosine gives
\begin{equation}
\begin{split}
\mathbf{A}(\mathbf{r}_\parallel, z, t) = {} & A_0 T_0 \big[ 1
+ 2\alpha\, e^{-G|z|} P(\mathbf{r}_\parallel) \big] \\
& \times \big( \cos\Omega t\, \hat{x} + \sin\Omega t\, \hat{y} \big) ,
\end{split}
\end{equation}
where $P(\mathbf{r}_\parallel) = \sum_{n=1}^{3} \cos(\mathbf{G}n \cdot \mathbf{r}_\parallel)$
is the hexagonal spatial profile. Evaluating at the graphene plane, $z = -z_0$, the decay factor becomes $e^{-G z_0}$, and it is
convenient to collect the coefficient of the pattern into a single modulation depth
$M(L) = 2\alpha, e^{-G z_0}$, which measures how strongly the superlattice is imprinted on the
light at the position of the sample. The transmitted vector potential at the graphene then
reads
\begin{equation}
\begin{split}
\mathbf{A}(\mathbf{r}_\parallel, t) = {} & A_0 T_0
\big[ 1 + M(L)\, P(\mathbf{r}_\parallel) \big] \\
& \times \big( \cos\Omega t\, \hat{x} + \sin\Omega t\, \hat{y} \big).
\end{split}\label{Eq:Aparallel}
\end{equation}
Inserting this field into the minimal substitution shifts the crystal wave vector according to
\begin{equation}
\mathbf{q}\to\mathbf{q}+\frac{e}{\hbar}\mathbf{A}(\mathbf{r}_\parallel t).
\end{equation}
The intralayer Dirac term therefore acquires the modulation
$a_0[1+M(L)P(\mathbf{r})]$, where
$a_0=(e/\hbar)A_0T_0$ is the driving strength of the circularly polarized field.

We next derive the vector potential that couples to the interlayer hopping - a description of the patterned waveguide light. Since the interlayer hopping occurs along the stacking direction of bilayer graphene, it requires a longitudinal component polarized along $\hat{\mathbf{z}}$~\cite{wang2016stacking}. Such a longitudinal component is forbidden in freely propagating light, whose vector potential is purely transverse. To overcome this limitation, we use a
rectangular waveguide oriented vertically as shown in Fig.\ref{case_2}. Light propagates in the waveguide from the top to the bottom exit, where it passes through a perforated plate and onto a bilayer graphene sheet at a distance $z=-z_0$. The guide has a rectangular cross-section
$a \times b$ in the $x$--$y$ plane, and the light propagates along $\hat{z}$. To determine the electromagnetic field inside the waveguide, we assume a time-harmonic vector
potential of the form
$\mathbf{A}(\mathbf{r},t)=\mathbf{A}_e(\mathbf{r})e^{-i\Omega t} $. As in the case of the circularly polarized field, we work in the Coulomb gauge, $\nabla\cdot\mathbf{A}=0$,
under which Maxwell's equations reduce to the Helmholtz equation for the spatial part of the
vector potential.
\begin{equation}
\big( \nabla^2 + k^2 \big) A_e = 0 , \qquad k = \Omega/c ,
\end{equation}
For a transverse-magnetic mode with $\mathbf{A}=A_z\hat{z}$, we separate the transverse profile from the propagation along the guide,
\begin{equation}
A_{e,z}(x,y,z)=u(x,y)\,\exp(-i k_z z),
\end{equation}
where $u(x,y)$ is the transverse mode profile and $k_z$ is the propagation constant. The conducting walls of the waveguide impose the boundary condition $A_z=0$ at $x=0,a$ and $y=0,b$. As a
result, the transverse field forms standing waves,
$u(x,y)=\sin\!\left(\frac{m\pi x}{a}\right)\sin\!\left(\frac{n\pi y}{b}\right)$, where $m,n\in\mathbb{Z}^+$. Substituting this form into the Helmholtz equation yields the transverse-magnetic eigenmodes
\begin{equation}
    A_z(x,y,z)=A_0\sin\!\left(\frac{m\pi x}{a}\right)\sin\!\left(\frac{n\pi y}{b}\right)\exp(-i k_z z),
\end{equation}
with
\begin{equation}
k_z = \sqrt{k^2-(m\pi/a)^2-(n\pi/b)^2}. 
\end{equation}
where $k_z$ (propagation constant) is the wavenumber in the z direction. The complete transverse-magnetic mode also contains in-plane components required by the Coulomb-gauge condition $\nabla\cdot\mathbf{A}=0$. Here, we focus on the longitudinal component of the lowest mode, $\mathrm{TM}_{11}$, corresponding to $m=n=1$. At the center of the waveguide cross-section, $x=a/2$ and $y=b/2$, the in-plane components vanish, while $A_z$ is maximal. For a real propagating mode, $k_z$ is real, which defines the cutoff frequency as
\begin{equation}
\Omega_c = \frac{1}{\sqrt{\mu\varepsilon}}
\sqrt{\left(\frac{m\pi}{a}\right)^2 + \left(\frac{n\pi}{b}\right)^2}.
\end{equation}
Thus, the longitudinal component of the vector potential at the waveguide exit, immediately before reaching the perforated plate, is
\begin{equation}
 A_z(r,t)=A_0\sin\!\left(\frac{m\pi x}{a}\right)\sin\!\left(\frac{n\pi y}{b}\right)\exp(-ik_z z-i\Omega{t})   
\end{equation}
A perforated plate spans the waveguide exit, and the bilayer graphene is centered with respect to the waveguide cross-section. At this position, the longitudinal mode profile is maximal and stationary. For a sample whose lateral dimensions are sufficiently small compared with $a$ and $b$, the profile varies only quadratically away from the center and can be absorbed into the amplitude. Setting the exit plane at $z=0$, the longitudinal component incident on the perforated plate is therefore approximated by
\begin{equation}
A_z(z=0,t)=A_0\cos(\Omega t),
\end{equation}
where $A_0$ is the amplitude of the longitudinal component at the center of the waveguide.

Within the scalar Kirchhoff approximation, we model the transmission of the longitudinal component using the same aperture function as for the in-plane field. We therefore write the field immediately below the plate as
$A_z(\mathbf{r}_\parallel,0,t)=T(\mathbf{r}_\parallel)A_0\cos\Omega t$.
Expanding the aperture function in reciprocal-lattice vectors gives
\begin{equation}
A_z(\mathbf{r}_\parallel, z, t) = A_0 T_0
\Big[1 + \frac{1}{T_0}\sum_{\mathbf{G}\neq 0} T_{\mathbf{G}}\,
e^{i\mathbf{G}\cdot\mathbf{r}_\parallel}\, e^{-|\mathbf{G}||z|}\Big]\cos\Omega t .
\end{equation}
Retaining the uniform term and the first shell, combining the six first-shell vectors into
cosines, and evaluating at the graphene plane, a distance $z_0$ below the plate, we obtain
\begin{equation}
A_z(\mathbf{r}, z_0, t) = A_0 T_0\big[1 + M(L) P(\mathbf{r})\big]\cos\Omega t ,
\label{Az}
\end{equation}
with the same hexagonal pattern $P(\mathbf{r})$, reciprocal magnitude $G = 4\pi/(\sqrt{3}L)$,
filling fraction $T_0$, contrast $\alpha$, and modulation depth $M(L) = 2\alpha e^{-G z_0}$ as in the circularly polarized case. A complete Maxwell solution for the perforated waveguide also contains transverse near-field components. In the electronic model considered here, we retain only the patterned component $A_z$ to isolate the effect of a spatially modulated interlayer hopping. The longitudinal drive should therefore be understood as an idealized effective driving protocol rather than as the complete electromagnetic field produced by the perforated waveguide.
We substitute  this field through the Peierls substitution, in which an interlayer hopping acquires
a phase set by the line integral of the vector potential along the path connecting the two
layers, $-i(e/\hbar)\int_{\mathbf{r}_i}^{\mathbf{r}_j}\mathbf{A}\cdot d\boldsymbol{\ell}$
~\cite{Peierls1933,Luttinger1951}. Since the two layers are separated only by the interlayer distance $d$, which is small
compared to the scale over which $\mathbf{A}$ varies, the field is essentially constant along
this path. The integral collapses to $A_z(\mathbf{r},t)\,d$. The interlayer hopping therefore
acquires the phase $\phi(\mathbf{r}, t) = (e/\hbar)A_z(\mathbf{r},t)\,d = \eta_0[1 +
M(L)P(\mathbf{r})]\cos\Omega t$, where $\eta_0 = (e/\hbar)A_0T_0 d$ is the dimensionless driving
strength of the longitudinal light. It plays the same role for the longitudinal drive that $a_0$
plays for the circularly polarized one, and we likewise treat it as a free parameter.

It is useful to relate the dimensionless driving strengths used in this work to experimentally accessible electromagnetic fields. The circularly polarized driving strength is defined as $a_0=(e/\hbar)A_0T_0$. For the largest value considered here, $a_0=0.30~\mathrm{nm}^{-1}$, the corresponding peak electric field is approximately $0.3\!-\!0.9~\mathrm{GV/m}$, depending on the driving frequency $\hbar\Omega=1\!-\!3~\mathrm{eV}$. This corresponds to intensities of order $10^{10}\!-\!10^{11}~\mathrm{W/cm^2}$, which are well within the range employed in Floquet experiments on graphene~\cite{McIver2020}. The longitudinal driving strength is characterized by the Peierls phase, $\eta_0=(e/\hbar)A_0T_0d$, where $d=0.34~\mathrm{nm}$ is the interlayer spacing. Because $d$ is much smaller than the in-plane length scale, achieving the largest value used in this work, $\eta_0=1.3$, requires stronger electric fields, approximately $4\!-\!11~\mathrm{GV/m}$, corresponding to intensities of order $10^{12}\!-\!10^{13}~\mathrm{W/cm^2}$. These values refer to the transmitted uniform-field amplitude at the graphene plane. Within the aperture model, the corresponding incident field is larger by a factor of $1/T_0$, while the incident intensity is larger by a factor of $1/T_0^2$. Although these fields are more demanding than those required for the circularly polarized drive, they remain experimentally achievable using intense few-cycle laser pulses and are comparable to those employed in high-harmonic-generation experiments on
graphene~\cite{Yoshikawa2017}.

\subsection{Model Hamiltonians}

We study the low-energy electronic states of bilayer graphene using the $k\cdot p$ effective Hamiltonian approach~\cite{McCann2013,McCann2006,Jung2011}, which more accurately describes the electronic structure near the $K$ and $K'$ points. We restrict our analysis to the two most common
stacking configurations, AA and AB (Bernal) stacked bilayer graphene. For each stacking, the low-energy Hamiltonian is constructed in the four-band basis
$(A_{1},B_{1},A_{2},B_{2})$, where the intralayer dynamics is described by the Dirac Hamiltonian and the stacking-dependent interlayer hopping couples the two layers.
\begin{figure*}
    \centering
\includegraphics[width=\linewidth]{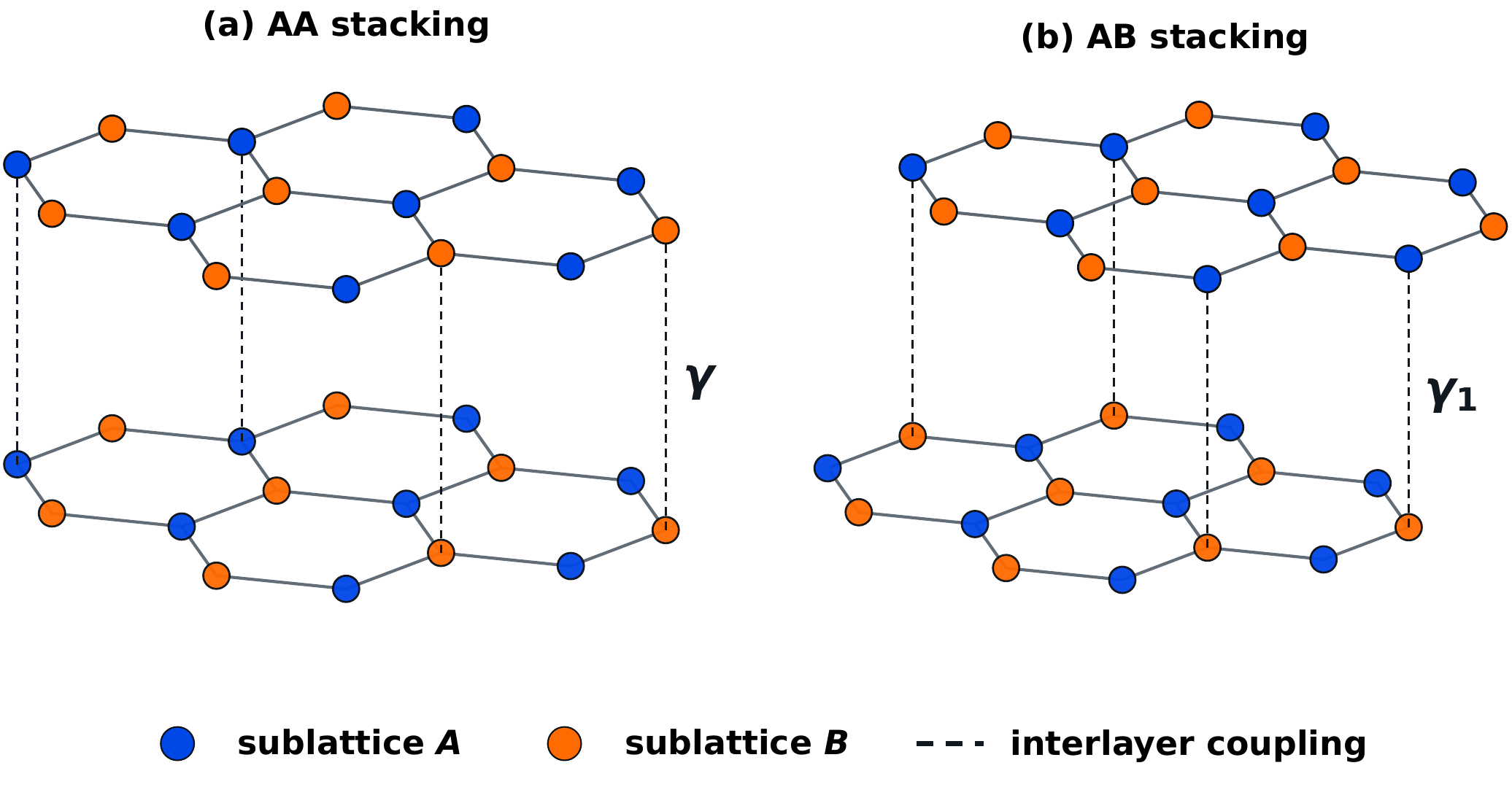}
    \caption{Stacking configurations of bilayer graphene. The $A$ and $B$ sublattices are
shown in blue and orange, respectively. The intralayer nearest-neighbor hopping is
denoted by $\gamma$ and vertical bonds represent the interlayer coupling.
(a)~In AA stacking, each atom lies directly above the same sublattice in the opposite
layer, giving interlayer hopping $\gamma$ between $A_{1}$--$A_{2}$ and
$B_{1}$--$B_{2}$. (b)~In AB (Bernal) stacking, the upper layer is shifted so that $A_{2}$
lies above $B_{1}$, forming the dimer pair coupled by $\gamma_{1}$, while $A_{1}$ and
$B_{2}$ remain non-dimer sites. }
    \label{fig:stacking}
\end{figure*}
We consider at a single valley $\xi=+1$ ($K$), with crystal momentum
$\mathbf{q}=(q_{x},q_{y})$ measured from the Dirac point, and define
$\pi= q_{x}+i q_{y}$. The topology is characterized by the valley Chern number
$C_{\xi}$, while the opposite valley ($K'$, $\xi=-1$) follows from time-reversal symmetry. In all that follows, we will loosely use the term Chern number even when we always mean valley Chern number.

For AA stacking, each atom in the upper layer lies directly above the corresponding
atom in the lower layer Fig.~\ref{fig:stacking}(a). The low-energy Hamiltonian in
the basis $(A_{1},B_{1},A_{2},B_{2})$ is~\cite{McCann2013,Lopes2007}
\begin{equation}
H_{\rm AA}=
\begin{pmatrix}
0 & \hbar v_{F}\pi^{\dagger} & \gamma & 0\\
\hbar v_{F}\pi & 0 & 0 & \gamma\\
\gamma & 0 & 0 & \hbar v_{F}\pi^{\dagger}\\
0 & \gamma & \hbar v_{F}\pi & 0
\end{pmatrix}.
\end{equation}
The terms $\hbar v_F \pi^{(\dagger)}$, with $\hbar v_F = 0.658$~eV$\cdot$nm (equivalently $v_{F}\approx1.0\times10^{6}~\mathrm{m/s}$) \cite{McCann2013}, describe the
intralayer Dirac coupling, while the interlayer hopping $\gamma = 0.20$~eV \cite{Tabert2012} couples the
same sublattices ($A_1$--$A_2$ and $B_1$--$B_2$). As a result, the two Dirac cones are shifted
to energies $\pm\gamma$, producing the spectrum $E = \pm\gamma \pm \hbar v_F |q|$.

For AB (Bernal) stacking, the upper layer is shifted so that the $A_{2}$ atoms lie
directly above the $B_{1}$ atoms, forming the dimer sites, while the remaining
sublattices are located above the centers of the hexagons
Fig.~\ref{fig:stacking}(b). The corresponding low-energy Hamiltonian is~\cite{McCann2013,McCann2006},
\begin{equation}
H_{\rm AB}=
\begin{pmatrix}
0 & \hbar v_{F}\pi^{\dagger} & 0 & 0\\
\hbar v_{F}\pi & 0 & \gamma_{1} & 0\\
0 & \gamma_{1} & 0 & \hbar v_{F}\pi^{\dagger}\\
0 & 0 & \hbar v_{F}\pi & 0
\end{pmatrix}.
\end{equation}
with $\hbar v_F = 0.658$~eV$\cdot$nm, equivalently $v_{F}\approx1.0\times10^{6}~\mathrm{m/s}$) and $\gamma_1 = 0.38$~eV~\cite{McCann2013}. Here $\gamma_1$  couples only the dimer sublattices $B_1$ and
$A_2$, while the non-dimer sublattices $A_1$ and $B_2$ remain uncoupled at this level of
approximation. Consequently, the dimer bands are split to energies $\pm\gamma_1$,
whereas the low-energy bands touch at zero energy with the quadratic dispersion
$E\simeq(\hbar v_{F}q)^{2}/\gamma_{1}$. Throughout this work, the weaker skew
interlayer hoppings $\gamma_{3}$ and $\gamma_{4}$ are neglected.

We now add the patterned circularly polarized and longitudinal light fields to the bilayer graphene Hamiltonian. The circularly polarized light enters the intralayer Dirac Hamiltonian through the minimal substitution $\pi\rightarrow\Pi=\pi+\tfrac{e}{\hbar}(A_x+iA_y)$  \cite{oka2009}.
Consequently, only the intralayer Dirac terms are modified, while the interlayer hopping
remains unchanged.

The longitudinal light couples to the interlayer hopping through the Peierls
substitution $\gamma^{(1)} \to \gamma^{(1)}
\exp[-i(e/\hbar)\int_{\mathbf{r}_i}^{\mathbf{r}_j} \mathbf{A}(\mathbf{r}, t) \cdot
d\boldsymbol{\ell}]$~\cite{Peierls1933, Luttinger1951}, where $\gamma^{(1)}$ denotes $\gamma$
for AA stacking and $\gamma_1$ for AB stacking. Here, the line integral runs along the interlayer hopping direction.
Since the vector potential points along $\hat{z}$ and the interlayer spacing
$d = 0.34$~nm~\cite{guinea2019} is much smaller than the length scale over which the field
varies; the field is essentially constant along the path, and the integral reduces to a
product, so that $\gamma^{(1)} \to \gamma^{(1)} e^{-i\phi(\mathbf{r}, t)}$ with Peierls phase
$\phi(\mathbf{r}, t) = (e/\hbar) A_z(\mathbf{r}, t)\, d$.
Thus, the longitudinal light modifies only the interlayer coupling, leaving the
intralayer Dirac Hamiltonian unchanged.
We now apply both substitutions; the time-dependent Hamiltonian for AA stacking is
\begin{equation}
H_{\rm AA}(\mathbf{r},t)=
\begin{pmatrix}
0 & \hbar v_{F}\Pi^{\dagger} & \gamma e^{-i\phi} & 0\\
\hbar v_{F}\Pi & 0 & 0 & \gamma e^{-i\phi}\\
\gamma e^{i\phi} & 0 & 0 & \hbar v_{F}\Pi^{\dagger}\\
0 & \gamma e^{i\phi} & \hbar v_{F}\Pi & 0
\end{pmatrix},
\label{eq:HAA_t}
\end{equation}
where the intralayer entries carry the shifted momentum $\Pi$ (from the circularly
polarized light) and the same-sublattice interlayer entries $\gamma$ carry the Peierls
phase $\phi$ (from the longitudinal light).
For AB stacking, the Hamiltonian is 
\begin{equation}
H_{\rm AB}(\mathbf{r},t)=
\begin{pmatrix}
0 & \hbar v_{F}\Pi^{\dagger} & 0 & 0\\
\hbar v_{F}\Pi & 0 & \gamma_{1}e^{-i\phi} & 0\\
0 & \gamma_{1}e^{i\phi} & 0 & \hbar v_{F}\Pi^{\dagger}\\
0 & 0 & \hbar v_{F}\Pi & 0
\end{pmatrix},
\label{eq:HAB_t}
\end{equation}
where now only the dimer coupling $\gamma_{1}$ (between $B_{1}$ and $A_{2}$) carries the
Peierls phase, while the intralayer entries again carry $\Pi$. In both cases, the patterned in-plane and longitudinal potentials
$\mathbf{A}_{\parallel}$ and $A_{z}$ Eqs.~\eqref{Eq:Aparallel} and \eqref{Az} share the
same spatial profile $1+M(L)P(\mathbf{r})$ and oscillate at frequency $\Omega$.

\subsection{Computational approach}
Since both Hamiltonians, Eqs.~\eqref{eq:HAA_t} and \eqref{eq:HAB_t}, are
periodic in space and time, we therefore apply Bloch's and Floquet's theorems 
simultaneously. 

Before introducing the Floquet formalism, we note that both driving fields are included in the Hamiltonian simultaneously. The circularly polarized light enters through the intralayer
minimal substitution, while the longitudinal light is incorporated through the Peierls phase in
the interlayer hopping. Different physical setups are obtained simply by choosing the driving
amplitudes. The patterned circularly polarized case corresponds to $a_0 \neq 0$ and
$\eta_0 = 0$, the longitudinal case to $a_0 = 0$ and $\eta_0 \neq 0$, and the combined case to
both $a_0 \neq 0$ and $\eta_0 \neq 0$. Recall also that, in the combined-drive setup, the circularly polarized field is taken to be uniform rather than patterned. This is
implemented by setting $M(L)=0$ in the circularly polarized field while retaining it in the
longitudinal one.  Using a single Hamiltonian for all three cases avoids
repeating the same derivation, while allowing the effects of each drive to be identified
separately. We now proceed to apply Bloch's and Floquet's theorems simultaneously.

Bloch's theorem associates the states with a crystal momentum 
$\mathbf{k}$ in the mini-Brillouin zone of the patterned superlattice, while 
Floquet's theorem assigns a quasienergy $\varepsilon$, defined modulo $\hbar\Omega$. 
The Floquet-Bloch states are therefore written as~\cite{Bukov2015,Oka2019},

\begin{equation}
\Psi_{\mathbf{k}}(\mathbf{r},t)=e^{i\mathbf{k}\cdot\mathbf{r}}\,
e^{-i\varepsilon t/\hbar}\,\Phi_{\mathbf{k}}(\mathbf{r},t),
\end{equation}
where $\Phi_{\mathbf{k}}$ is periodic in both space and time,
$\Phi_{\mathbf{k}}(\mathbf{r}+\mathbf{R},t)=\Phi_{\mathbf{k}}(\mathbf{r},t)$ and
$\Phi_{\mathbf{k}}(\mathbf{r},t+T)=\Phi_{\mathbf{k}}(\mathbf{r},t)$. 
We then substitute $\Psi_{\mathbf{k}}(\mathbf{r},t)$ into the time-dependent Schr\"odinger equation
$H(\mathbf{r},t)\,\Psi_{\mathbf{k}}=i\hbar\,\partial_{t}\Psi_{\mathbf{k}}$ and obtain the so-called Floquet-Bloch equation
\begin{equation}
\label{eq:FB}
\big[H(\hat{\mathbf{q}}+\mathbf{k},\mathbf{r},t)-i\hbar\,\partial_{t}\big]\,
\Phi_{\mathbf{k}}(\mathbf{r},t)=\varepsilon\,\Phi_{\mathbf{k}}(\mathbf{r},t),
\end{equation}
Since $\Phi_{\mathbf{k}}$ is periodic in $\mathbf{r}$ and $t$, we expand it in plane waves over the
reciprocal lattice $\{\mathbf{G}\}$ of the pattern and in Fourier harmonics of the
drive.

\begin{equation}
\label{eq:expand}
\Phi_{\mathbf{k}}(\mathbf{r},t)=\sum_{\mathbf{G},\,l}
c^{\mathbf{k}}_{\mathbf{G},l}\,
e^{i\mathbf{G}\cdot\mathbf{r}}\,e^{-il\Omega t},
\end{equation}
Similarly, we can  expand the Hamiltonian,
$H(\mathbf{k},\mathbf{r},t)=\sum_{\mathbf{G},l}
H^{(l)}_{\mathbf{G}}(\mathbf{k})\,e^{i\mathbf{G}\cdot\mathbf{r}}\,e^{-il\Omega t}$.
The static Hamiltonian is diagonal in both the photon index and the reciprocal lattice vector, and it is the drives that generate the harmonics of $H$; their oscillation in time couples
different photon sectors, while the patterned part of each drive couples different reciprocal
lattice vectors. The circularly polarized field
enters the intralayer blocks through $\Pi$ and, being harmonic in time, generates only
the static ($l=0$) Dirac term and the $l=\pm1$ sidebands. The longitudinal field
enters the interlayer coupling as $\gamma^{(1)}e^{-i\phi}$ with
$\phi(\mathbf{r},t)=\eta(\mathbf{r})\cos\Omega t$ and
$\eta(\mathbf{r})=\eta_{0}\!\left[1+M(L)P(\mathbf{r})\right]$; using the Jacobi--Anger identity~\cite{NISTDLMF},
\begin{equation}
\label{eq:JA}
e^{-i\eta(\mathbf{r})\cos\Omega t}
=\sum_{l=-\infty}^{\infty}(-i)^{l}\,J_{l}\!\big(\eta(\mathbf{r})\big)\,e^{-il\Omega t},
\end{equation}
so the interlayer coupling therefore contains all Floquet harmonics $l$, with each harmonic weighted by the corresponding Bessel function $J_{l}$
of the first kind. Its plane-wave expansion is
\begin{equation}
\big[J_{l}(\eta)\big]_{\mathbf{G}}
=\frac{1}{A_{\rm uc}}\int_{\rm uc}
J_{l}\!\big(\eta(\mathbf{r})\big)\,e^{-i\mathbf{G}\cdot\mathbf{r}}\,d^{2}r,
\label{eq:JlG}
\end{equation}
with $A_{\rm uc}$ the area of the superlattice unit cell. We match  Fourier components in Eq.~\eqref{eq:FB} and get a
time-independent eigenvalue problem in frequency (Sambe) space. The central equation is 
\begin{equation}
\label{eq:central}
\sum_{\mathbf{G}',\,l'}
\mathcal{H}^{\,l-l'}_{\mathbf{G}\mathbf{G}'}(\mathbf{k})\,
c^{\mathbf{k}}_{\mathbf{G}',l'}
+\,l\hbar\Omega\,c^{\mathbf{k}}_{\mathbf{G},l}
=\varepsilon\,c^{\mathbf{k}}_{\mathbf{G},l}\; 
\end{equation}
where $l\hbar\Omega$ is the photon-energy ladder and the block
$\mathcal{H}^{\,l-l'}_{\mathbf{G}\mathbf{G}'}$ connects photon sectors
$l'\!\to\!l$ and reciprocal-lattice vectors $\mathbf{G}'\!\to\!\mathbf{G}$. In the
four-band basis $(A_{1},B_{1},A_{2},B_{2})$, the block takes the following form for both stackings
\begin{equation}
\label{eq:block}
\mathcal{H}^{\,l-l'}_{\mathbf{G}\mathbf{G}'}(\mathbf{k})=
\begin{pmatrix}
h^{(l-l')}_{\mathbf{G}-\mathbf{G}'}(\mathbf{k}+\mathbf{G}') &
\;\mathcal{T}^{(l-l')}_{\mathbf{G}-\mathbf{G}'}\\[6pt]
\big[\mathcal{T}^{(l'-l)}_{\mathbf{G}'-\mathbf{G}}\big]^{\dagger} &
\;h^{(l-l')}_{\mathbf{G}-\mathbf{G}'}(\mathbf{k}+\mathbf{G}')
\end{pmatrix},
\end{equation}
with intralayer and interlayer $2\times2$ blocks
\begin{align}
h^{(0)}_{\mathbf{0}}(\mathbf{k}+\mathbf{G}) &=
\hbar v_{F}
\begin{pmatrix} 0 & \pi^{\dagger}_{\mathbf{k}+\mathbf{G}}\\
\pi_{\mathbf{k}+\mathbf{G}} & 0 \end{pmatrix},\\[4pt]
h^{(+1)}_{\mathbf{G}} &=
\hbar v_{F}\,[a]_{\mathbf{G}}
\begin{pmatrix} 0 & 0\\ 1 & 0 \end{pmatrix},\label{sidebands+}\\[4pt]
h^{(-1)}_{\mathbf{G}} &=
\hbar v_{F}\,[a]_{\mathbf{G}}
\begin{pmatrix} 0 & 1\\ 0 & 0 \end{pmatrix},\label{sidebands-}\\[4pt]
\mathcal{T}^{(l)}_{\mathbf{G}} &=
\mathcal{T}_{0}\,(-i)^{l}\,\big[J_{l}(\eta)\big]_{\mathbf{G}}.
\end{align}
where the shifted intralayer momenta are (at valley $\xi=+1$)
\begin{equation}
\label{eq:pidef}
\begin{aligned}
\pi_{\mathbf{k}+\mathbf{G}} &= (k_x+G_x) + i(k_y+G_y),\\
\pi^{\dagger}_{\mathbf{k}+\mathbf{G}} &= (k_x+G_x) - i(k_y+G_y),
\end{aligned}
\end{equation}
Here $[a]_{\mathbf{G}}$ are the Fourier components of the in-plane drive amplitude
$a(\mathbf{r}) = a_0 [1 + M(L) P(\mathbf{r})]$. Because we truncated the transmitted field to
the first shell, the pattern function
$P(\mathbf{r}) = \sum_{n=1}^{3} \cos(\mathbf{G}_n \cdot \mathbf{r})$ contains only the six
first-shell vectors, and each cosine splits into two exponentials with weight $1/2$, so the
expansion terminates and
\begin{equation}
[a]_{\mathbf{G}} =
\begin{cases}
a_0 , & \mathbf{G} = 0 , \\[4pt]
\dfrac{a_0 M(L)}{2} , & \mathbf{G} \in \{ \pm\mathbf{G}_1, \pm\mathbf{G}_2, \pm\mathbf{G}_3 \} , \\[4pt]
0 , & \text{otherwise} .
\end{cases}
\end{equation} 
The components $[J_l(\eta)]_{\mathbf{G}}$ of the longitudinal drive (see Eq.\eqref{eq:JlG} and notice the additional integral over position) cannot be written in closed
form. Although $P(\mathbf{r})$ contains only the first shell, it appears inside the Bessel
function, so $J_l(\eta_0 [1 + M(L) P(\mathbf{r})])$ has nonvanishing components at every shell.
We obtain them numerically by evaluating the Bessel function on a real-space grid over the
superlattice unit cell and taking its discrete Fourier transform, retaining components up to the
shell truncation used. 
The in-plane drive
amplitude $a(\mathbf{r})=a_{0}[1+M(L)P(\mathbf{r})]$ enters through its Fourier
components $[a]_{\mathbf{G}}$. The stacking enters \emph{only} through the constant
interlayer matrix $\mathcal{T}_{0}$,
\begin{equation}
\label{eq:T0}
\mathcal{T}_{0}^{\rm AA}=\gamma\,\mathbb{1}_{2},
\qquad
\mathcal{T}_{0}^{\rm AB}=
\begin{pmatrix} 0 & 0\\ \gamma_{1} & 0 \end{pmatrix},
\end{equation}
so that Eqs.~\eqref{eq:central}--\eqref{eq:T0} describe AA and AB stacking on the same
footing. In the absence of the longitudinal drive ($\eta_{0}=0$)
$J_{l}(\eta)\to\delta_{l0}$ and $\mathcal{T}^{(l)}_{\mathbf{G}}\to
\mathcal{T}_{0}\,\delta_{l0}\delta_{\mathbf{G}\mathbf{0}}$, that recover the static
interlayer coupling and in the absence of the circular drive ($a_{0}=0$) the sidebands
$h^{(\pm1)}$ vanish.

For numerical calculations, the Floquet--Bloch Hamiltonian in Sambe space is truncated to a
finite number of reciprocal-lattice shells and Floquet replicas. The reciprocal-lattice vectors
are grouped into shells, where each shell contains vectors related by the point-group symmetry
of the superlattice and therefore having the same magnitude. For a hexagonal lattice, the first
shell consists of the six vectors $\pm\mathbf{G}_{1,2,3}$. The truncation is performed shell by
shell rather than by selecting individual reciprocal-lattice vectors, ensuring that all
symmetry-related vectors are retained together. This preserves the symmetry of the full
Hamiltonian and avoids introducing artificial anisotropy into the quasienergy spectrum.
Convergence with respect to both the reciprocal-lattice and Floquet truncations is verified
explicitly~\cite{PhysRevA.7.2203}. With $N_s$ shells of
reciprocal-lattice vectors and $N_F$ Floquet copies with four bands per reciprocal vector, the dimension of our truncated matrix is  $4\,N_G\,N_F$.


Next, we diagonalize the truncated matrix to determine eigenvalues and the corresponding eigenvectors. The quasienergy spectrum obtained from the diagonalization determines the Floquet band structure. In the high-frequency regime, to characterize the topological properties of the isolated quasienergy bands, the band dispersion alone is insufficient.

A topological invariant is an interesting quantity to evaluate for isolated bands. In our case, we consider Chern numbers, which are computed using the gauge-invariant lattice formulation introduced by Fukui \textit{et al.}~\cite{Fukui2005}. The mini-Brillouin zone is discretized into a uniform mesh, and the occupied Floquet eigenvectors are calculated at every momentum point. From the overlap between neighboring eigenvectors, link variables are constructed, from which the lattice Berry curvature is obtained on each plaquette. The Chern number is computed with the method of Fukui, Hatsugai and Suzuki~\cite{Fukui2005},
in which the mini-Brillouin zone is discretized into a grid of plaquettes. For each pair of
neighboring grid points, we define the link variable
\begin{equation}
U_\mu(\mathbf{k}) = \frac{\langle u_n(\mathbf{k}) | u_n(\mathbf{k} + \Delta \mathbf{k}\mu) \rangle}
{\big| \langle u_n(\mathbf{k}) | u_n(\mathbf{k} + \Delta \mathbf{k}\mu) \rangle \big|} ,
\qquad \mu = x, y ,
\end{equation}
where $|u_n(\mathbf{k})\rangle$ is the periodic part of the Bloch state of band $n$ and
$\Delta \mathbf{k}\mu$ is the grid spacing along $\mu$. The link variable is a pure phase, and
it measures the phase acquired in moving from one grid point to the next. Taking the product of
the four link variables around a plaquette gives the Berry curvature on that plaquette, and
summing over the whole mini-Brillouin zone gives the Chern number
\begin{equation}
C = \frac{1}{2\pi i} \sum{\mathbf{k}} \ln \Big[
U_x(\mathbf{k}) U_y(\mathbf{k} + \Delta \mathbf{k}_x)
U_x^{-1}(\mathbf{k} + \Delta \mathbf{k}_y) U_y^{-1}(\mathbf{k}) \Big] ,
\end{equation}
with the branch of the logarithm chosen in $(-\pi, \pi]$. Because each link variable is
normalized to unit modulus, any arbitrary phase of the Bloch states cancels between neighboring
links, so the method is gauge invariant and yields integer Chern numbers on a discretized
Brillouin zone.

\subsection{Convergence}
Formally, the Floquet--Bloch Hamiltonian is infinite because it contains both infinitely many
reciprocal-lattice vectors and infinitely many Floquet harmonics. In numerical calculations,
both must therefore be truncated. The first truncation is in the number of reciprocal-lattice
shells, denoted by $N_s$, and the second is in the number of Floquet replicas, denoted by
$N_F$. Keeping $N_s$ shells includes a total of $N_G(N_s)$ reciprocal-lattice vectors. Since
each reciprocal-lattice vector is combined with $N_F$ Floquet replicas and the four electronic
states of the bilayer (two layers and two sublattices), the resulting Floquet--Bloch
Hamiltonian has the dimension
\begin{equation}
D = 4\, N_G(N_s)\, N_F .
\end{equation}
To check convergence, we increase each truncation in turn and measure how much the bands of
interest move. For the shell truncation we define
\begin{equation}
\Delta_{N_s} = \max_{\mathbf{k}, n}
\big| \varepsilon_n^{(N_s + 2)}(\mathbf{k}) - \varepsilon_n^{(N_s)}(\mathbf{k}) \big| ,
\end{equation}
where the maximum runs over the central bands $n$ and over the momenta $\mathbf{k}$ along the
high-symmetry path, and $\Delta_{N_F}$ is defined in the same way for the Floquet truncation. We
perform this check for every parameter set used in this work and retain only results for which
both $\Delta_{N_s}$ and $\Delta_{N_F}$ fall below $10^{-3}$~meV. This is far below the width of
the flat bands (that we discover later), which is of order a few tenths of a meV, so the results are converged on the
scale of interest. The truncation required depends on the strength of the patterned part of the drive, since only
that part couples different reciprocal-lattice vectors, and on the superlattice period, since
the modulation depth $M(L)$ grows with $L$. The largest truncation needed to meet this criterion
is $N_s = 5$, corresponding to $N_G = 37$ reciprocal-lattice vectors, together with $N_F = 5$
Floquet replicas, giving a matrix of dimension $D = 740$.

\section{Results}
\label{results}
\subsection{Band structure results}
In this section, we investigate the quasienergy band structures of both AA- and AB-stacked bilayer graphene under spatially patterned periodic driving. We consider six driving configurations: (i) AA bilayer with patterned circularly polarized light, (ii) AA bilayer with patterned longitudinal light, (iii) AA bilayer with patterned longitudinal light together with uniform circularly polarized light, (iv) AB bilayer with patterned circularly polarized light, (v) AB bilayer with patterned longitudinal light, and (vi) AB bilayer with patterned longitudinal light together with uniform circularly polarized light. We calculate the corresponding quasienergy spectra along the high-symmetry path of the mini-Brillouin zone using the converged Floquet–Bloch truncations established in the previous section. By comparing these six cases, we examine how the stacking geometry and the nature of the periodic drive influence the low-energy band structure.\

\begin{figure*}
\centering
\includegraphics[width=\linewidth]{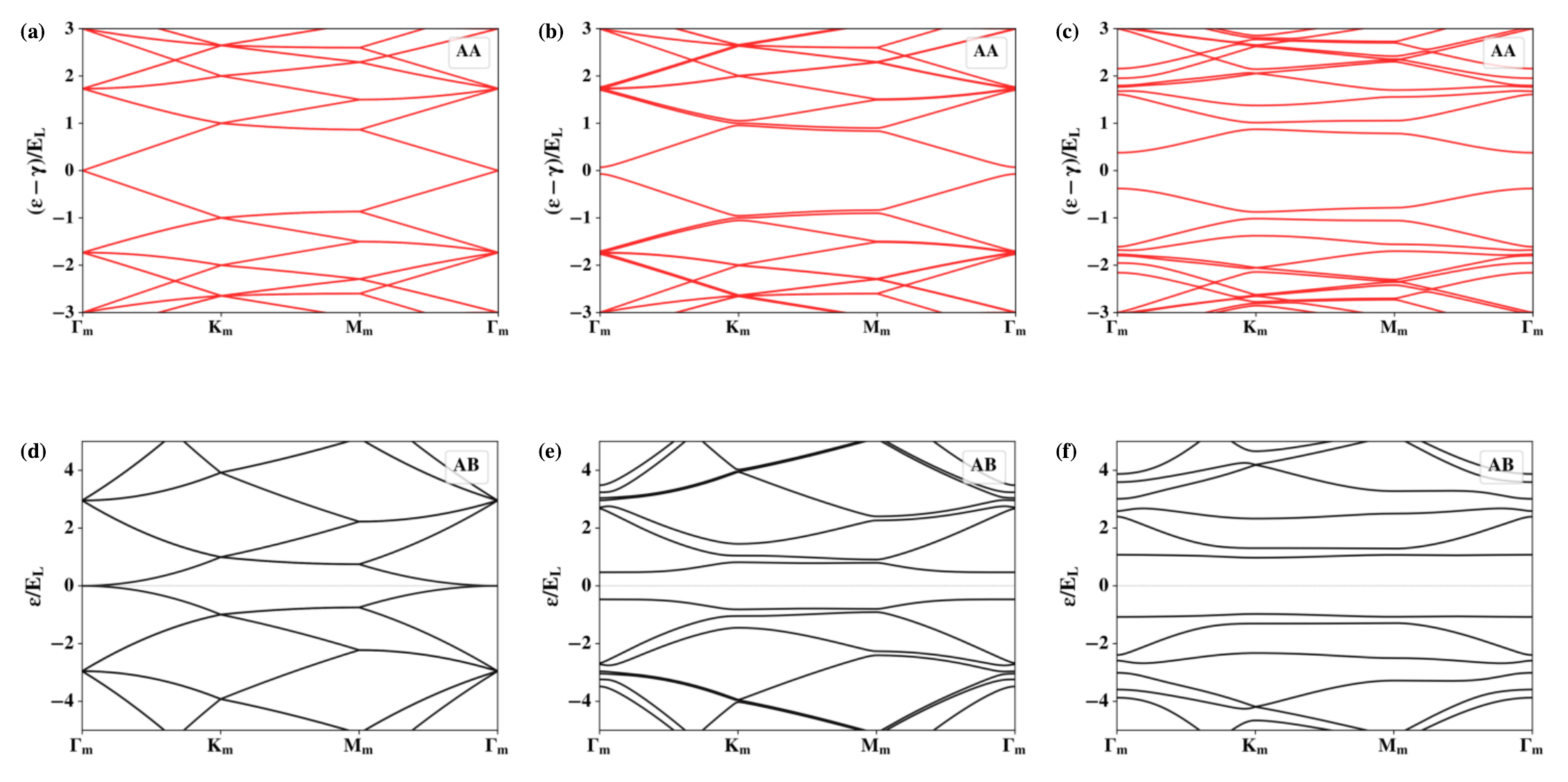}
\caption{Evolution of the quasienergy band structure under increasing patterned
circularly polarized drive. Panels (a)–(c) show AA-stacked bilayer graphene for
$a_0=0.00$, $0.10$, and $0.25$, respectively; panels (d)–(f) show AB-stacked
bilayer graphene for $a_0=0.00$, $0.07$, and $0.12$. In all cases, $L=100$~nm and
$\hbar\Omega=3$~eV. The quasienergy is plotted in units of $E_L$, defined separately
for the two stackings as described in the text: for AA we show
$(\varepsilon-\gamma)/E_L$ with $E_L=\hbar v_F|K_m|$, and for AB we show
$\varepsilon/E_L$ with $E_L=(\hbar v_F|K_m|)^2/\gamma_1$. }

\label{CPL}
\end{figure*}

We first examine the effect of patterned circularly polarized light on the quasienergy band structures of AA- and AB-stacked bilayer graphene. To understand how the Floquet spectrum evolves with the strength of the driving field, we calculate the band structures for increasing values of the driving amplitude while keeping the superlattice period ($L=100~\mathrm{nm}$) and driving frequency ($\hbar\Omega=3~\mathrm{eV}$) fixed. The resulting quasienergy band structures are shown in Fig.~\ref{CPL}. For AA stacking Figs.~\ref{CPL}(a)–(c), increasing the drive strength progressively renormalizes the low-energy bands. As the driving amplitude increases, the central quasienergy band becomes progressively isolated from the surrounding minibands, while a small gap opens at the Dirac point. In contrast, the AB bilayer Figs.~\ref{CPL}(d)–(f) exhibits a stronger reconstruction of the low-energy spectrum. As the driving amplitude increases, the central quadratic bands become increasingly distorted, the minibands are displaced in energy, and the gaps between neighboring bands are noticeably modified. As a result, the AB bilayer develops a well-isolated and relatively flat central Floquet band at substantially smaller driving amplitudes. In contrast, the AA bilayer requires much stronger driving to achieve comparable band isolation, and the resulting central Floquet band remains more dispersive than in the AB case. These features closely mimic behavior known from conventional moir\'e materials under
circularly polarized light, where the drive is known to flatten the low-energy bands and to
open gaps at the Dirac points, and where the interplay between the moir\'e potential and the
drive controls the isolation of the central band~\cite{Topp2019,Li2020}. It is
interesting that we recover the same qualitative behavior here without any twist, using only
a patterned circularly polarized electromagnetic field, which suggests that the light-induced superlattice plays a
role closely analogous to the moir\'e potential.

\begin{figure*}
\centering
\includegraphics[width=1\linewidth]{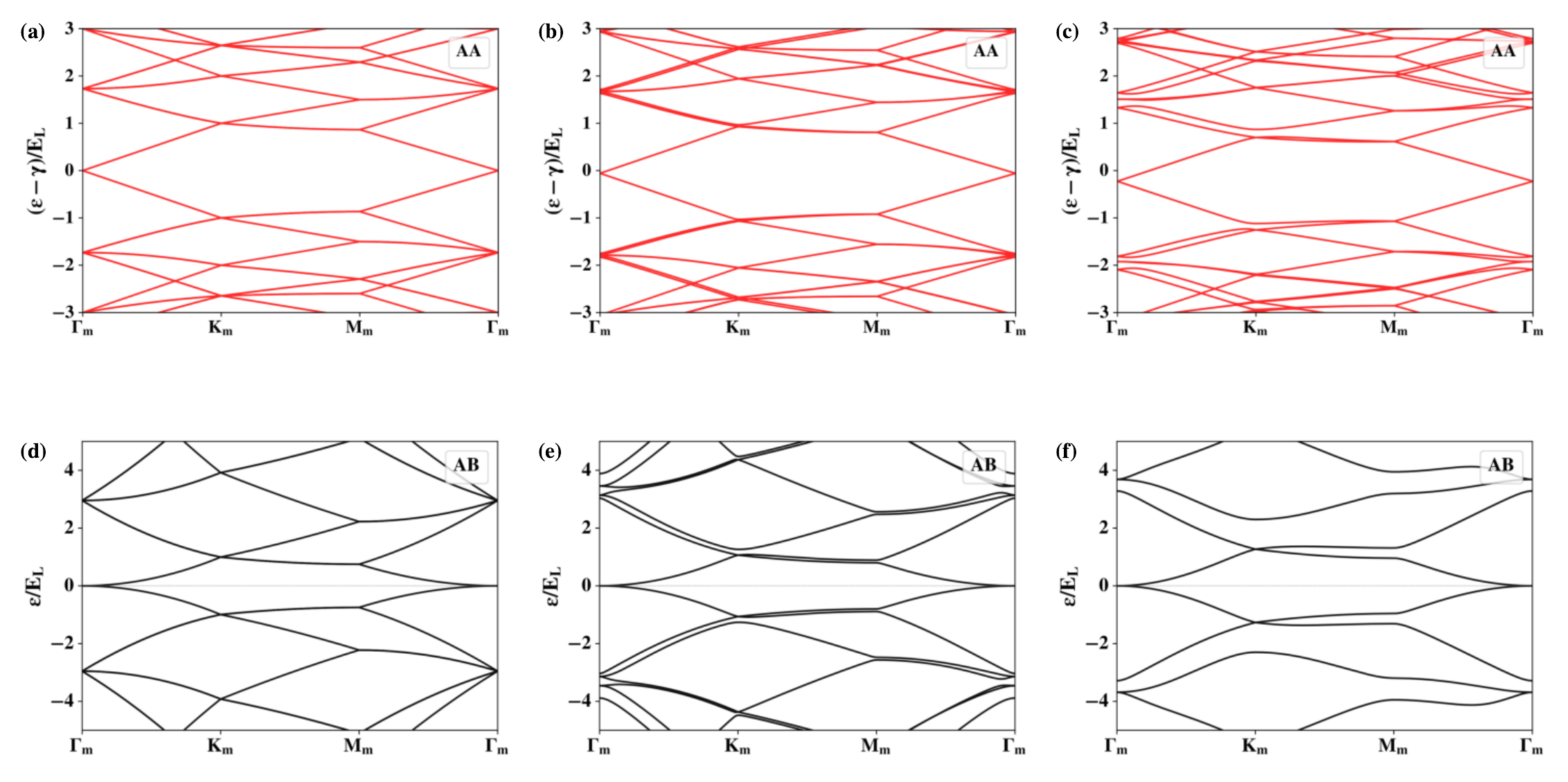}
\caption{Evolution of the quasienergy band structure under increasing patterned
longitudinal drive. Panels (a)–(c) show AA-stacked bilayer graphene for $\eta_0=0.0$, $0.15$, and $0.30$, respectively; panels (d)–(f) show AB-stacked
bilayer graphene for $\eta_0=0.0$, $0.6$, and $1.2$. In all cases, $L=100$~nm and
$\hbar\Omega=3$~eV. The quasienergy is plotted in units of $E_L$, defined
separately for the two stackings as described in the text.}
\label{LOG}
\end{figure*}

We next investigate the effect of the patterned longitudinal drive, which naively is the configuration expected to most closely mimic the physics of undriven twisted bilayer graphene, since it directly modulates the interlayer coupling in a spatially periodic manner. We calculate the band structure for increasing driving amplitudes, keeping the superlattice period
$L=100$~nm and drive frequency $\hbar\Omega=3$~eV fixed. The resulting
quasienergy band structures are shown in Fig.~\ref{LOG}, for AA stacking with
$\eta_0=0.0$, $0.15$, and $0.30$ Figs.~\ref{LOG}(a)–(c) and for AB stacking with
$\eta_0=0.0$, $0.6$, and $0.12$ Figs.~\ref{LOG}(d)–(f). We find that the patterned longitudinal drive preserves time-reversal symmetry; therefore, no gap opens at the Dirac points for either AA or AB stacking. The band crossings remain intact over the entire range of driving amplitudes considered. Instead, the drive mainly renormalizes the interlayer coupling. For AA stacking,
this appears as a downward shift of the Dirac cones. For AB stacking, the same renormalization reduces the effective dimer coupling, which increases the curvature and bandwidth of the central quadratic bands. In addition, the spatial pattern opens avoided-crossing gaps between the folded minibands, but only in the parts of the mini-Brillouin zone allowed by symmetry. A gap opens at the $M_m$ point while
the band touchings at the $K_m$ point remain protected, as shown in Fig.~\ref{LOG}(e)-(f). The Dirac point itself stays gapless for both stackings, as required by time-reversal symmetry. As a
result, the central bands remain connected to the surrounding spectrum, and no isolated Floquet band forms in either stacking. It is natural to ask why a drive that directly modulates the interlayer coupling still does not replicate the physics of twisted bilayer graphene. The key reason is that the longitudinal drive changes only the magnitude of the interlayer coupling, not its structure. In twisted bilayer graphene, the local stacking varies continuously across the moir\'e unit cell, producing AA, AB, and intermediate stacking regions. This spatial variation changes how the sublattices in the two layers are connected and is responsible for opening gaps at the Dirac crossings and isolating the flat bands. In contrast, the longitudinal drive multiplies a fixed interlayer hopping matrix by a position-dependent scalar. It therefore strengthens or weakens the interlayer coupling from one region to another while preserving the same sublattice connectivity throughout the system.

\begin{figure*}
\centering
\includegraphics[width=\linewidth]{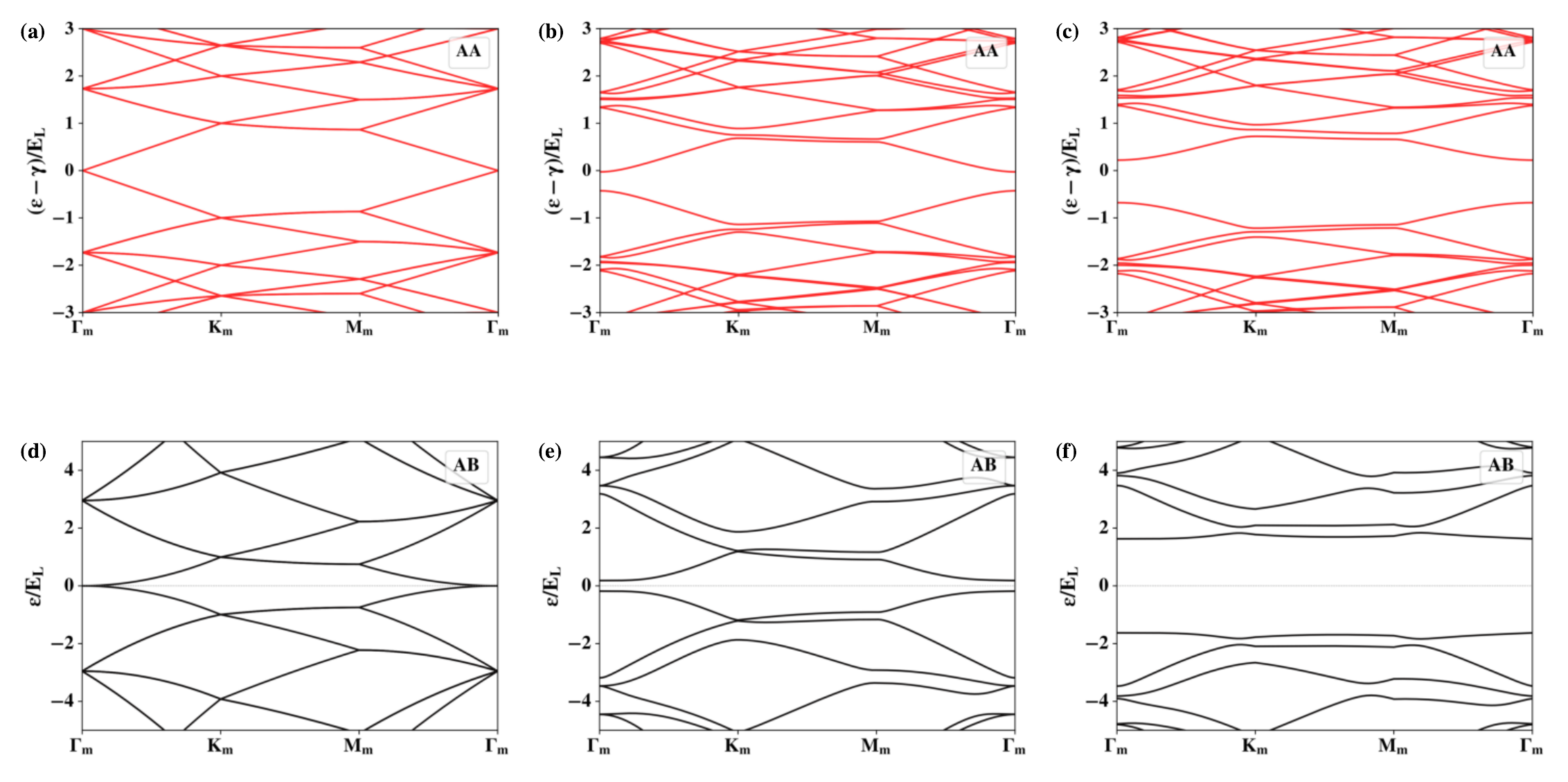}
\caption{Quasienergy band structures under the combined drive: patterned
longitudinal light plus uniform circularly polarized light. Panels (a)- (c) show
AA-stacked bilayer graphene at $\eta_0=0.0$, $a_0=0.0$, and at  fixed $\eta_0=0.3$ for $a_0=0.20$ and  $0.30$,  respectively, and panels (d)-(f) show AB-stacked bilayer graphene at $\eta_0=0.0$, $a_0=0.0$, and at fixed $\eta_0=1.0$ for $a_0=0.05$ and  $0.15$.  In all cases, $L=100$~nm and
$\hbar\Omega=3$~eV. The quasienergy is plotted in units of $E_L$.}
\label{LOG_CPL}
\end{figure*}

The results of the previous section show that a patterned longitudinal drive alone creates a superlattice but does not isolate the central Floquet bands, because it preserves the symmetry protecting the Dirac crossings. This naturally raises the question of whether band isolation can be achieved by introducing the missing symmetry breaking. To address this, we combine the patterned longitudinal drive with a uniform circularly polarized field. It is interesting to see that under the patterned longitudinal and uniform circularly polarized drive, at fixed
superlattice period $L=100$ nm and drive frequency $\hbar\Omega=3$ eV, shown in Fig.~\ref{LOG_CPL}, a gap opens at the Dirac points for both stackings, which did not happen
for the longitudinal drive alone. The gap increases with the circularly polarized amplitude. For the AB bilayer, Figs.~\ref{LOG_CPL}(d)-(f) show that the central bands also flatten as the amplitude increases, and they completely separate from the other bands, leaving an isolated pair of nearly flat bands around
zero energy. In the AA bilayer, a gap opens at the cone tips once the circularly polarized part of the drive is added, and this gap grows with amplitude, but the bands coming from the cones stay dispersive even at an amplitude nearly twice as large as the one used for AB. In
this sense, the AA bilayer behaves very differently from the AB bilayer. The combined drive therefore produces flat isolated gapped bands in the AB bilayer,
while in the AA bilayer it opens a gap but the bands remain dispersive. This configuration provides the closest analog in our model to a driven moir\'e material. In twisted bilayer graphene, circularly polarized light is known to gap the Dirac crossings and isolate the flat bands. Interestingly, the AB bilayer exhibits the same behavior here without any twist, suggesting that the essential ingredients are not the moir\'e structure itself but the combination of a spatially periodic modulation and broken time-reversal symmetry. An important advantage of our approach is that we control these two ingredients independently. The mask determines the superlattice period, the longitudinal drive controls the strength of the periodic modulation, and the circularly polarized field sets the size of the quasienergy gap. All three parameters can therefore be tuned in situ rather than being fixed during fabrication. However, the AA bilayer never reaches this regime, even under substantially stronger driving. This shows that the Floquet response depends not only on the external driving field but also on the underlying electronic band structure. In particular, the quadratic band touching in the AB bilayer is more easily reconstructed than the symmetry-protected Dirac cones in the AA bilayer.

Before discussing a fixed driving strength while varying other parameters, it is instructive to compare the effects of the three driving schemes. The most effective of them is the patterned circularly polarized drive. At the lowest driving amplitudes, it enters the intralayer Dirac Hamiltonian via the minimal substitution while simultaneously breaking time-reversal symmetry, generating flat and well-isolated central Floquet bands. Consequently, it creates the superlattice and the quasienergy gap in a single driving field. In contrast, the patterned longitudinal drive is the least effective. It generates a pronounced superlattice, but only modulates the magnitude of the interlayer coupling, leaving the Dirac crossings protected and preventing the formation of isolated bands. The combined drive overcomes this limitation by adding the missing time-reversal symmetry breaking via the uniform circularly polarized field. Therefore, it also creates isolated flat bands in the AB bilayer, but requires two independent driving fields instead of one. This comparison leads to two observations. First, for both the patterned circularly polarized and the combined drives, flat isolated bands are generated in the AB bilayer.

In contrast, for the patterned longitudinal drive only, they are never formed. This demonstrates that the essential element to isolate the Floquet bands is the breaking of time-reversal symmetry, not the enhancement of the strength of the periodic modulation. Second, the AA bilayer does not develop flat bands like AB under any of the three driving schemes, even at significantly larger driving strengths. The Dirac cones of AA stacking can be gapped and shifted but remain very dispersive. By contrast, the quadratic band touching of the AB bilayer is readily rebuilt into narrow and well-isolated Floquet bands. These results illustrate that stacking geometry is crucial for how the electronic structure responds to periodic driving.

So far we have varied only the driving strength, keeping the period and frequency fixed. The
natural next questions are what happens when the frequency is lowered or when the superlattice
period is made longer, keeping the driving strength fixed. We find that lowering the driving frequency or increasing the superlattice period produces band structures that are qualitatively similar to those obtained by increasing the driving strength. For this reason, we present the corresponding results in the Appendix.  We note only that both parameter sweeps lead to progressively flatter and more isolated central Floquet bands in the AB bilayer. In contrast, in the AA bilayer the Dirac crossings become gapped, but the bands remain dispersive.

\subsection{Effective Hamiltonians and explanation of bandstructure results}

In the previous section, we examined the quasienergy band structures produced by the different
driving protocols, each of which led to a distinct physical outcome. The patterned circularly
polarized drive opened gaps at the Dirac points and isolated a nearly flat central band in AB, much
like circularly polarized light in a moir\'e superlattice. In contrast, the patterned
longitudinal drive preserved the Dirac crossings and merely reshaped the bands, leaving the
system closer to an undriven twisted bilayer than to a driven one. Combining the longitudinal
drive with a uniform circularly polarized field restored both the gap and the isolated flat
bands in AB and dispersive bands in AA. Although these features are evident in the numerical spectra, the spectra alone do not
reveal their physical origin. In particular, they do not show which terms generated by the
driving fields are responsible for opening the gap, creating the superlattice, or flattening
the bands.

To answer these questions, we construct an effective static Hamiltonian that reproduces the
periodically driven system in the high-frequency limit. The underlying idea is analogous to the
Kapitza pendulum\cite{Kapitza1951,Bukov2015}, where a rapidly oscillating pivot can effectively be replaced by a static effective
potential that even stabilizes the originally unstable inverted pendulum position. In the same spirit, the Van Vleck
expansion replaces the time-periodic bilayer Hamiltonian by a time-independent effective
Hamiltonian, organized as a series in inverse driving frequency, whose individual terms can be identified and directly connected to the physical effects observed in the quasienergy spectra.

The Van Vleck expansion to lowest order\cite{Bukov2015,Goldman2014,Eckardt2015} gives the effective Hamiltonian as
\begin{equation}
H_{\rm eff} = H^{(0)} + \sum_{l \geq 1}
\frac{\big[ H^{(l)}, H^{(-l)} \big]}{l\,\hbar\Omega} + \mathcal{O}(\Omega^{-2}) ,
\end{equation}
where $H^{(l)}$ are the Fourier components of the time-periodic Hamiltonian over one driving period
\begin{equation}
H^{(l)} = \frac{1}{T}\int_{0}^{T} H(t)\, e^{il\Omega t}\, dt .
\end{equation}
Each term can be built from the
intralayer and interlayer harmonics of Eq.~\eqref{sidebands+},~\eqref{sidebands-} as
$H^{(l)} = h^{(l)}(\mathbf{k}+\mathbf{G}) + T^{(l)}_{\mathbf{G}}$. The zeroth component $H^{(0)}$
is the time-averaged Hamiltonian, and $H^{(\pm 1)}$ additional information from the drive. The leading
correction is a commutator of the harmonics, so a drive generates a new term only when its
harmonics do not commute.

We begin with the patterned circularly polarized drive, which produced flat isolated bands in the AB bilayer but only opened a gap in the AA bilayer. For this drive, the
longitudinal field is absent, so the interlayer harmonics vanish,
$T^{(l)}_{\mathbf{G}} = T^{(0)}_{\mathbf{G}}\delta_{l,0}$, and the only nonzero harmonics are the
static block $h^{(0)}_{\mathbf{G}}(\mathbf{k}+\mathbf{G})$ and the two sidebands
$h^{(\pm 1)}_{\mathbf{G}}$. The first-order Van Vleck correction is therefore the single
commutator $[h^{(+1)}, h^{(-1)}]/\hbar\Omega$. Using the expressions in Eq.~\eqref{sidebands+} and~\eqref{sidebands-}, it acts
within each layer as
\begin{equation}
\big[ h^{(+1)}, h^{(-1)} \big] = \big(\hbar v_F [a]_{\mathbf{G}}\big)^2
\begin{pmatrix} -1 & 0 \\ 0 & 1 \end{pmatrix} ,
\end{equation}
so the effective Hamiltonian acquires a term proportional to $-\sigma_z$ in each layer,

\begin{equation}
H_{\rm eff} = \mathbb{1}\otimes h^{(0)}(\mathbf{k}+\mathbf{G}) + \mathcal{T}
- \frac{(\hbar v_F\, a(\mathbf{r}_\parallel))^2}{\hbar\Omega}\,\mathbb{1}\otimes\sigma_z ,
\end{equation}
Here $\sigma_i$ are Pauli matrices and $\mathcal{T}$ is the
$4\times4$ interlayer coupling built from the $2\times2$ block $\mathcal{T}_0$, corresponding to
$\mathcal{T}_0^{\rm AB}$ for AB stacking and $\mathcal{T}_0^{\rm AA}$ for AA stacking. The last term, generated by the drive, is a Haldane-type mass (although position-dependent), with opposite signs on the two
sublattices of each layer, familiar from graphene irradiated by circularly polarized light. Its
the handedness of the light determines the sign; it breaks time-reversal symmetry, and it
opens the gap at the Dirac points as shown in Fig.~\ref{CPL}. Since the circularly polarized field couples only to the
intralayer Dirac term, while the difference between AA and AB stacking resides entirely in the
static interlayer matrix $\mathcal{T}_0$, which does not enter the commutator that generates the mass,
this mass is identical for both stackings. The distinction between AA and AB therefore enters
only through the static interlayer coupling, which is $\gamma$ between like sublattices
($A_1$--$A_2$ and $B_1$--$B_2$) for AA stacking and the dimer coupling $\gamma_1$ on the
$B_1$--$A_2$ bond for AB stacking.
The mass is spatially modulated because the driving amplitude inherits the pattern of the mask,
$a(\mathbf{r}_\parallel) = a_0[1 + M(L) P(\mathbf{r}_\parallel)]$, and enters the effective
Hamiltonian quadratically. The resulting mass,
\begin{equation}
m(\mathbf{r}_\parallel) = \frac{(\hbar v_F a_0)^2}{\hbar\Omega}
\big[1 + M(L) P(\mathbf{r}_\parallel)\big]^2 ,
\end{equation}
is therefore periodic on the superlattice scale. The patterned circularly polarized drive thus
provides both ingredients required to form isolated flat bands. The Haldane mass (constant term) already breaks
time-reversal symmetry and gaps the Dirac crossings, while its spatial modulation generates the
superlattice that folds and flattens the bands. This explains why the patterned circularly
polarized drive alone produces an isolated flat band in the AB bilayer Figs.~\ref{CPL} (d)-(f). In the AA bilayer,
however, the starting point is two rigid Dirac cones shifted to $\pm\gamma$ rather than a
quadratic band touching at zero energy, so the same mass opens a gap at the cone tips but cannot
reorganize the spectrum into an isolated flat-band manifold Figs.~\ref{CPL}(a)-(c).\\

We next consider the patterned longitudinal drive alone, which naively is the configuration expected
to mimic the physics of undriven twisted bilayer graphene. Here, however, for both stackings, in the quasienergy plots, the Dirac crossings were left intact, and only the band details were slightly modified. Notice in particular that for this drive the
circularly polarized field is absent and
$h^{(\pm 1)}_{\mathbf{G}} = 0$. The only harmonics that survive are the interlayer terms
$T^{(l)}_{\mathbf{G}} = \mathcal{T}_0\,(-i)^l [J_l(\eta)]_{\mathbf{G}}$, present at every $l$ through the
Bessel expansion of the Peierls phase.
The first-order Van Vleck correction requires the commutator $[H^{(l)}, H^{(-l)}]$ for $l \geq 1$,
where now $H^{(l)} = T^{(l)}$. Each harmonic is a scalar Bessel factor multiplying the same fixed
interlayer matrix $\mathcal{T}_0$ , so
\begin{equation}
\big[ H^{(l)}, H^{(-l)} \big] = (-i)^l (i)^l [J_l(\eta)]_{\mathbf{G}} [J_{-l}(\eta)]_{\mathbf{G}}
\big[ T_0, T_0 \big] = 0 ,
\end{equation}
for both stackings, since any matrix commutes with itself. The first-order correction vanishes
identically, and the effective Hamiltonian is simply the time average,
\begin{equation}
H_{\rm eff} = \mathbb{1}\otimes h^{(0)}(\mathbf{k}+\mathbf{G}) + \mathcal{T}\, J_0\big(\eta(\mathbf{r}_\parallel)\big) .
\end{equation}
The drive generates no new term in the effective Hamiltonian. Its only effect is to renormalize
the interlayer coupling through the Bessel factor $J_0[\eta(\mathbf{r}_\parallel)]$, which
rescales $\mathcal{T}_0$ (with a weak position dependence). Consequently, no
Haldane mass is generated, time-reversal symmetry remains intact, and the Dirac crossings stay
protected regardless of the driving strength, in agreement with the numerical spectra. The
renormalization $J_0[\eta(\mathbf{r}_\parallel)]$ is itself spatially periodic and therefore
creates a superlattice. However, a periodic modulation of a single scalar coupling, without
breaking time-reversal symmetry, cannot gap the Dirac crossings or isolate a flat band. As a
result, the patterned longitudinal drive shifts the AA Dirac cones. It broadens the
central bands of the AB bilayer as shown in Fig.~\ref{LOG}, without producing the isolated flat bands obtained under the
patterned circularly polarized drive.\\

Since the longitudinal drive alone does not open a gap, we now add a uniform circularly polarized
field to supply the Haldane mass. Both drives are present, so each harmonic contains an
intralayer piece from the circularly polarized light and an interlayer piece from the
longitudinal light,
\begin{equation}
H^{(\pm 1)} = h^{(\pm 1)} + T^{(\pm 1)} ,
\end{equation}
and the commutator that enters the Van Vleck expansion splits into four pieces,
\begin{equation}
\begin{split}
\big[ H^{(1)}, H^{(-1)} \big] = {} & \big[ h^{(1)}, h^{(-1)} \big]
+ \big[ h^{(1)}, T^{(-1)} \big] \\
& + \big[ T^{(1)}, h^{(-1)} \big] + \big[ T^{(1)}, T^{(-1)} \big] .
\end{split}
\end{equation}
The first commutator is the Haldane mass already found for the circularly polarized drive above. The
last term is zero because both harmonics are proportional to the same
interlayer matrix in the longitudinal drive case. The two middle commutators are the cross terms, which mix the intralayer
sidebands with the interlayer harmonics, and they contain everything new generated by the combined drive. Evaluating the cross term, we find that it vanishes identically for AA stacking but is nonzero
for AB stacking, where it generates a new interlayer coupling,
\begin{equation}
\frac{i\,\hbar v_F\, a\, \gamma_1 J_1\big(\eta(\mathbf{r}_\parallel)\big)}{\hbar\Omega} (\sigma_x\otimes\sigma_z)
\end{equation}
This term couples the layers on the $A_1$--$A_2$ and $B_1$--$B_2$ bonds with
opposite signs on the two sublattices, and is absent from the static AB
Hamiltonian, whose only interlayer bond is the $B_1$--$A_2$ dimer. Because this
coupling is odd under sublattice exchange, it lowers the symmetry of the AB
bilayer relative to its undriven form. These are precisely the same-sublattice
interlayer bonds ($A_1$--$A_2$, $B_1$--$B_2$) that AA stacking possesses at rest,
which is why the cross term vanishes for AA. The drive has nothing new to
generate there. The combined drive thus dynamically endows the AB bilayer with
an interlayer connection that AA has intrinsically.
The effect of this cross term on the flat bands is negligible, however. The gap is set by the
Haldane mass and the isolation by the superlattice, and the cross term only slightly modifies the
curvature of the central bands while acting mainly on the high-energy dimer manifold. It
nonetheless represents a genuine structural difference between the stackings, since the same term vanishes identically for AA. The essential distinction between the two stackings therefore lies
in their starting point. Under the combined drive, the AB bilayer, with its soft quadratic
touching, is reorganized into an isolated flat manifold, as shown in Fig.~\ref{LOG_CPL}(a)-(c), while the AA
bilayer, whose rigid cones are shifted to $\pm\gamma$, is only gapped and shifted,
Fig.~\ref{LOG_CPL}(d)-(f). This is why the combined drive isolates flat bands in the AB bilayer but not in the AA bilayer.

\subsection{Results for band topology}

The quasienergy spectra presented in the previous section naturally raise the question of which of the driven Floquet bands are topologically nontrivial. A Chern number can be assigned to any band that is spectrally isolated from the rest of the spectrum, irrespective of whether the band is flat or dispersive. As shown previously, both AA- and AB-stacked bilayer graphene develop isolated bands, although the two stackings exhibit markedly different behavior. In AB stacking, the central bands become isolated and nearly flat at relatively small driving strengths. In AA stacking, the central bands also isolate at small driving strengths, but the gaps separating them from the surrounding bands are very small, and even at stronger driving these bands remain dispersive.

Nevertheless, in both stackings the four central bands are sufficiently isolated that their Chern numbers are well defined. We have verified numerically that bands beyond the four central ones can also be spectrally isolated and carry well-defined Chern numbers. Here, however, we restrict the analysis to the four central bands because they govern the low-energy physics near charge neutrality and display the most relevant drive-induced features. A systematic topological classification of the remaining Floquet bands lies beyond the scope of the present work. We verified the reported Chern numbers to be converged with respect to the $k$-space discretization by repeating the calculations on progressively finer meshes.

As discussed in the previous section, spectral isolation occurs only for two of the three driving configurations considered here: patterned circularly polarized light and the combined drive consisting of patterned longitudinal light together with uniform circularly polarized light. Our topological analysis is therefore restricted to these two cases, where we determine the Chern numbers of the four central bands for both AA and AB stacking. The patterned longitudinal drive alone is excluded because, as shown in Sec.~\ref{results}, it preserves the Dirac crossings, preventing the central bands from becoming spectrally isolated and therefore precluding the definition of a Chern number.

\begin{figure*}
    \centering
    \includegraphics[width=\linewidth]{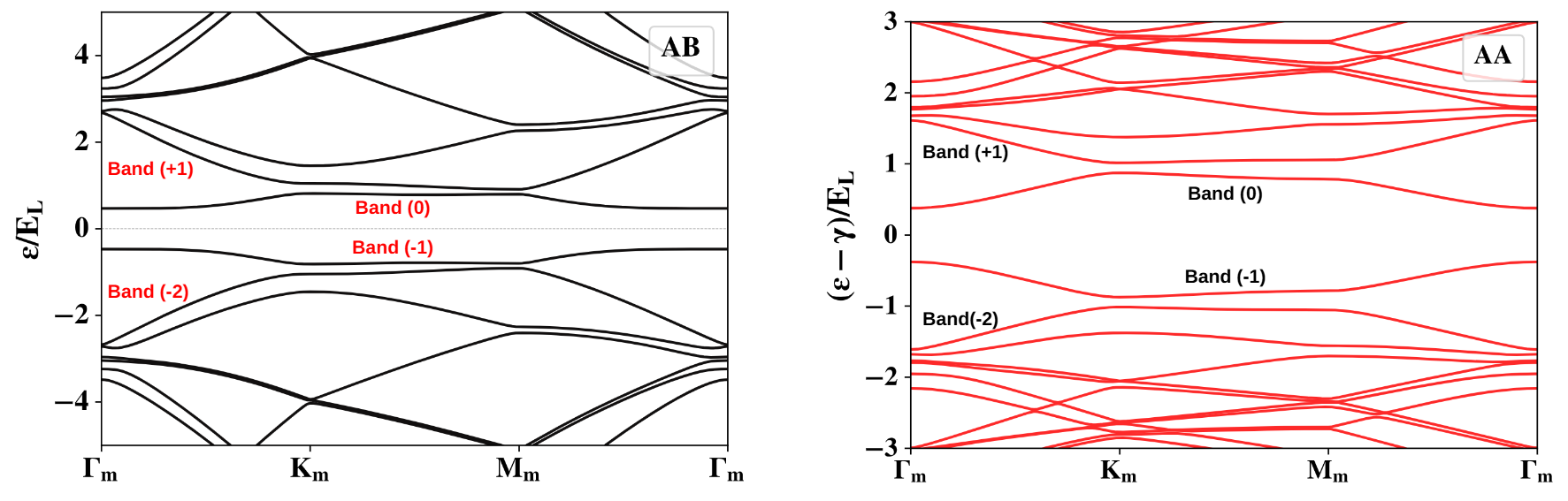}
    \caption{The four central bands, labeled band $(-2)$ through band $(+1)$
    from lowest to highest, for (a) AB and (b) AA stacked bilayer graphene.}
    \label{Numbering}
\end{figure*}

In this section, we construct the topological phase diagrams by computing
the Chern numbers of the four central bands at a fixed driving frequency
over the parameter space spanned by the driving strength and the
superlattice period. The four central
bands considered throughout in the phase diagrams are labeled from the lowest, band $(-2)$,
through bands $(-1)$ and $(0)$, to the highest, band $(+1)$, as shown in
Fig.~\ref{Numbering} for both the AB and AA stackings. This numbering is
used consistently in all phase diagrams and Chern-number assignments. The Chern numbers are obtained from the Floquet
effective Hamiltonian derived in the high-frequency (Van~Vleck) expansion,
rather than from the full Floquet operator. We checked numerically that
this effective description is consistent with the full diagonalization,
reproducing the same Chern numbers across the parameter space at a small fraction of the computational cost. Separate phase diagrams are presented
for each stacking and each driving configuration. To illustrate the mechanism underlying the topological transitions, we also examine a representative example showing the evolution of the band energies together with their corresponding Chern numbers. This allows us to identify each topological transition with a closing and reopening of the energy gap between adjacent bands, accompanied by an exchange of Chern number between them. Finally, we note that although the quasienergy spectra obtained in the frequency and superlattice-period sweeps presented in the Appendix closely resemble those discussed in the main text, their topological character is not necessarily identical, since the Chern number is determined by the eigenvectors of the Floquet Hamiltonian rather than by the quasienergy spectrum alone.

\begin{figure*}
    \centering    \includegraphics[width=1\linewidth]{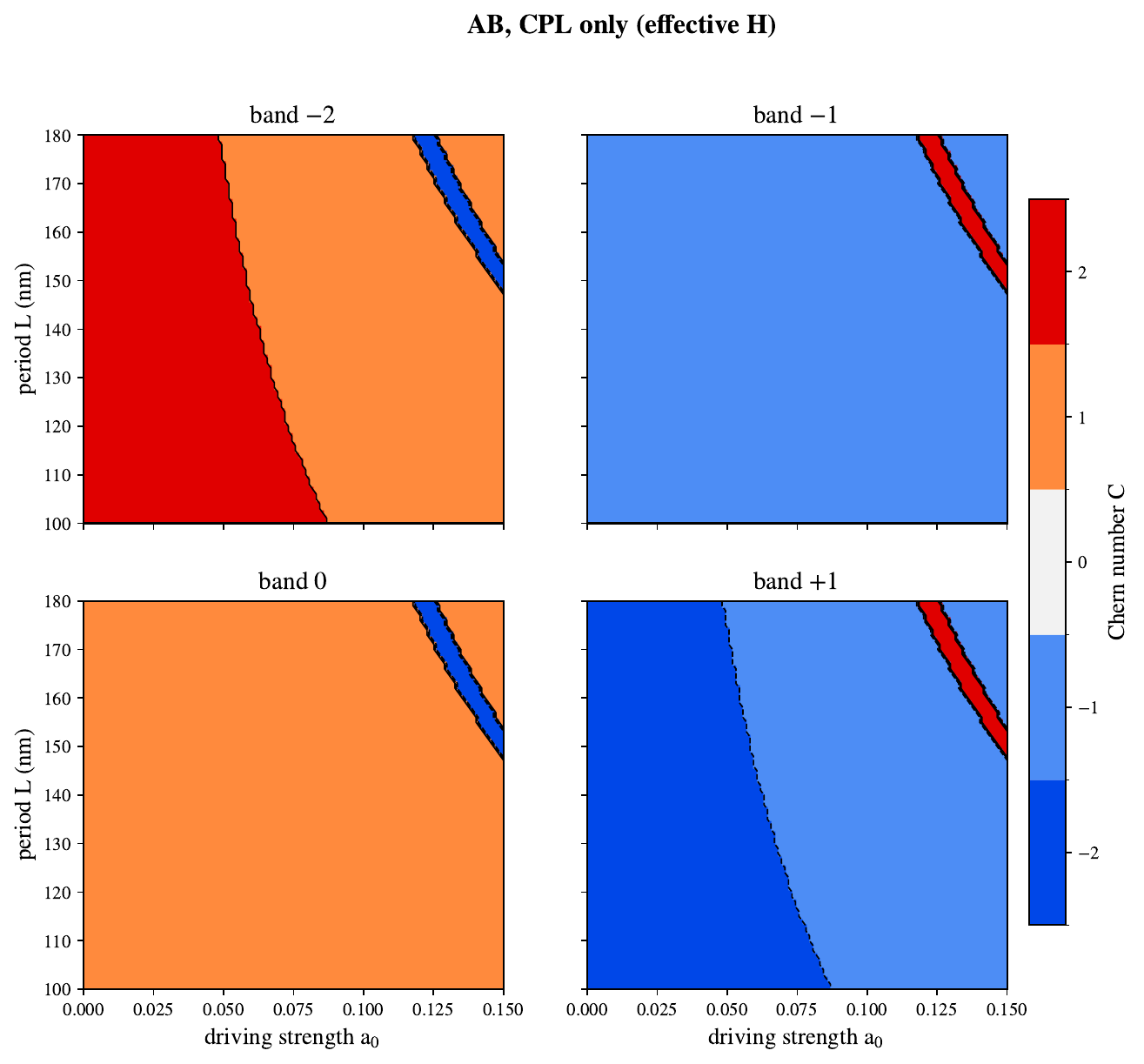}
    \caption{Phase diagram of the four central bands of the AB bilayer under the
patterned circularly polarized drive, showing the Chern number over the plane of
driving strength $a_0 \in [0,0.15]$ and superlattice period $L \in [100,180]$~nm
at fixed frequency $\hbar\Omega = 3$~eV. The four panels show the central bands
ordered by energy from the lowest band to the highest band. Bands
$-1$ and $0$ are the two inner bands and bands $-2$ and $+1$ the two outer bands.
Colored regions give the integer Chern number where the band is spectrally
isolated over the full Brillouin zone.}
    \label{CPL_density}
\end{figure*}

\begin{figure*}
    \centering
    \includegraphics[width=1\linewidth]{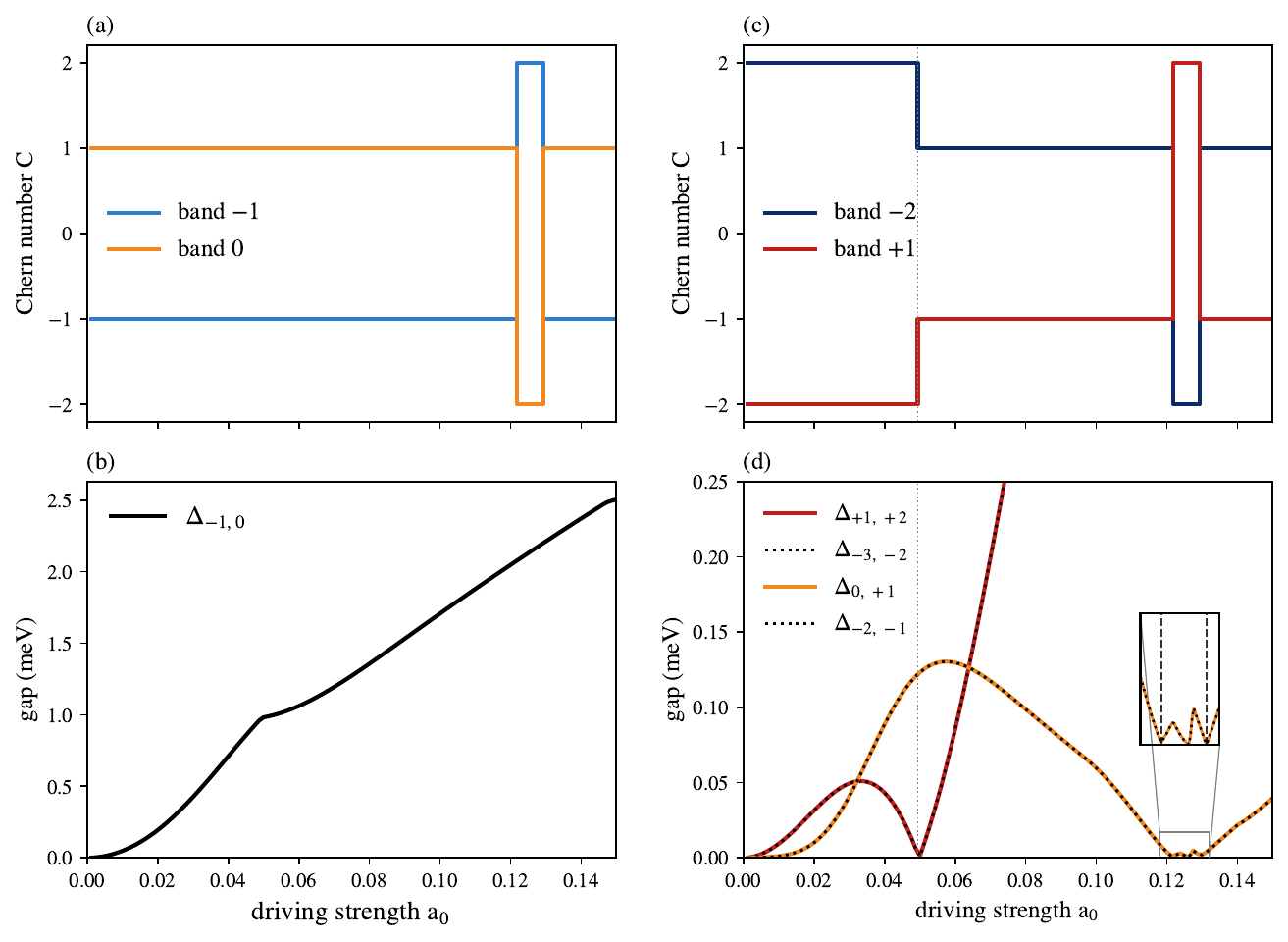}
    \caption{
  Line cut of the phase diagram Fig~\ref{CPL_density} at fixed period $L = 175$~nm.
  (a),(c)~Chern numbers of the inner bands $(-1,0)$ and the outer bands
  $(-2,+1)$ versus driving strength $a_0$. (b),(d)~Corresponding minimum direct
  gaps. Vertical dotted lines mark the transitions, and the inset in (d)
  magnifies the re-entrant window near $a_0 \approx 0.12$--$0.13$, to confirm gap closing. Between the two gap
  closings that bound the window, there is a third gap closing at which the Chern
  numbers do not change.
}
    \label{linecute_AB}
\end{figure*}

\begin{figure*}
    \centering
    \includegraphics[width=\linewidth]{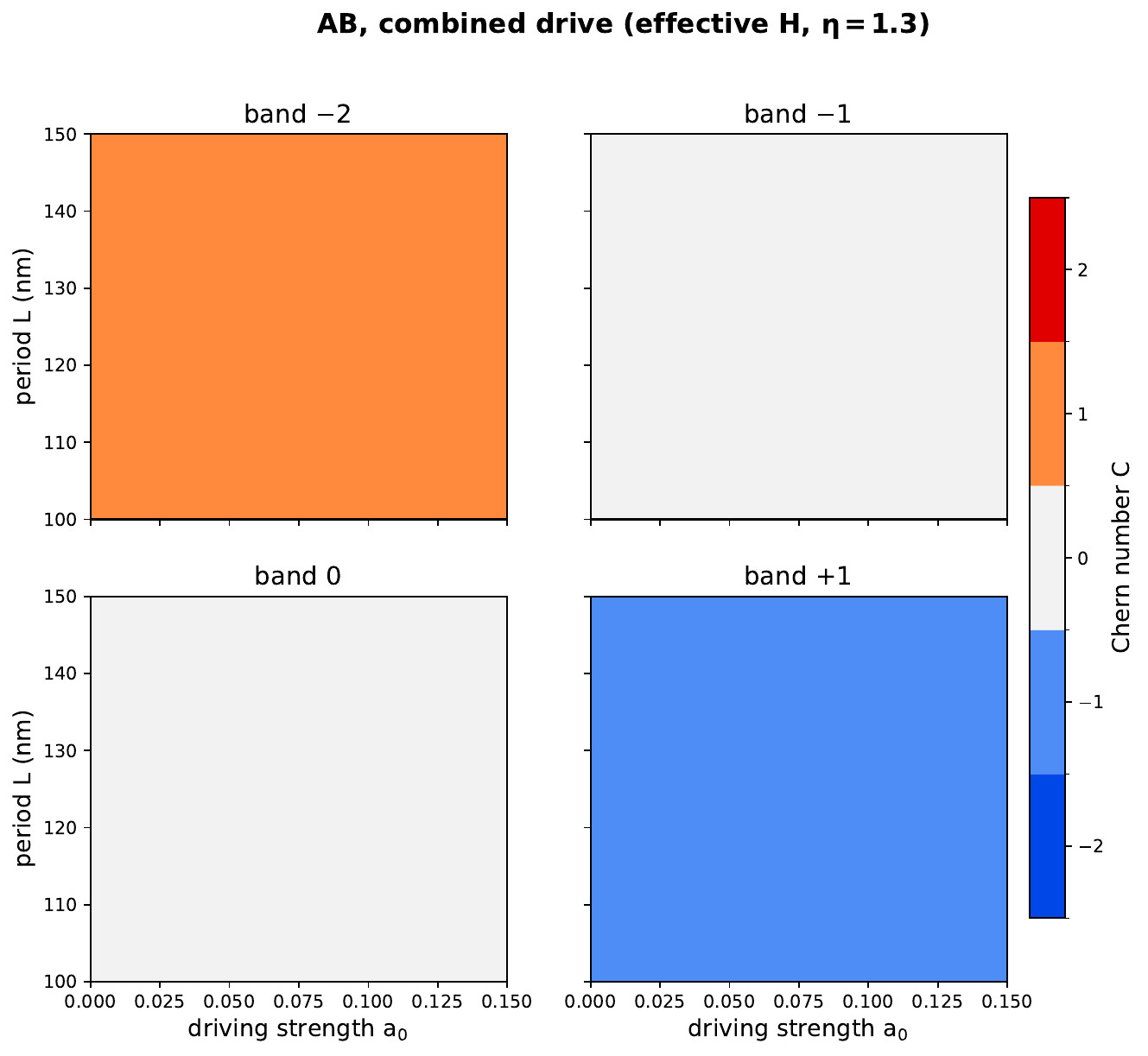}
    \caption{Phase diagrams of the four central bands of AB-stacked bilayer graphene
under the combined drive. The Chern numbers are shown as functions of the circularly
polarized driving strength $a_0 \in [0,0.15]$ and the superlattice period
$L \in [100,150]$~nm, with the longitudinal driving strength fixed at
$\eta_0 = 1.3$ and the driving frequency fixed at $\hbar\Omega = 3$~eV. The band
ordering, labeling, and color coding are the same as in Fig.~\ref{CPL_density}.}
    \label{phase_AB_combined}
\end{figure*}

\begin{figure*}
    \centering
    \includegraphics[width=\linewidth]{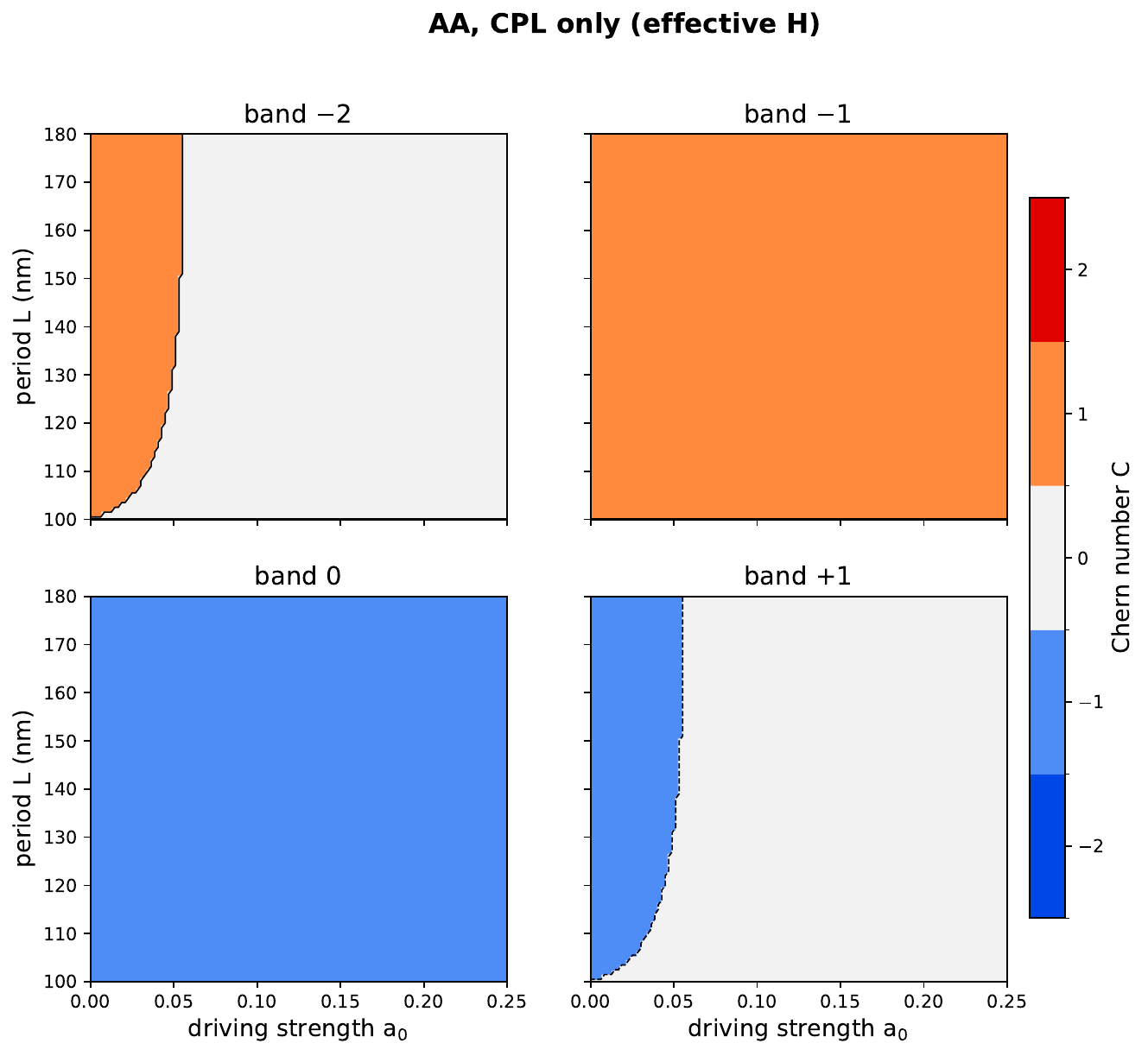}
    \caption{Phase diagrams of the four central bands of AA-stacked bilayer graphene
under the patterned circularly polarized drive. The Chern numbers are shown as
functions of the driving strength $a_0 \in [0,0.25]$ and the superlattice period
$L \in [100,180]$~nm, at fixed frequency $\hbar\Omega = 3$~eV. The band ordering,
labeling, and color coding are the same as in Fig.~\ref{CPL_density}.}
    \label{phase_diagram_AA_CPL}
\end{figure*}

\begin{figure*}
    \centering
    \includegraphics[width=1\linewidth]{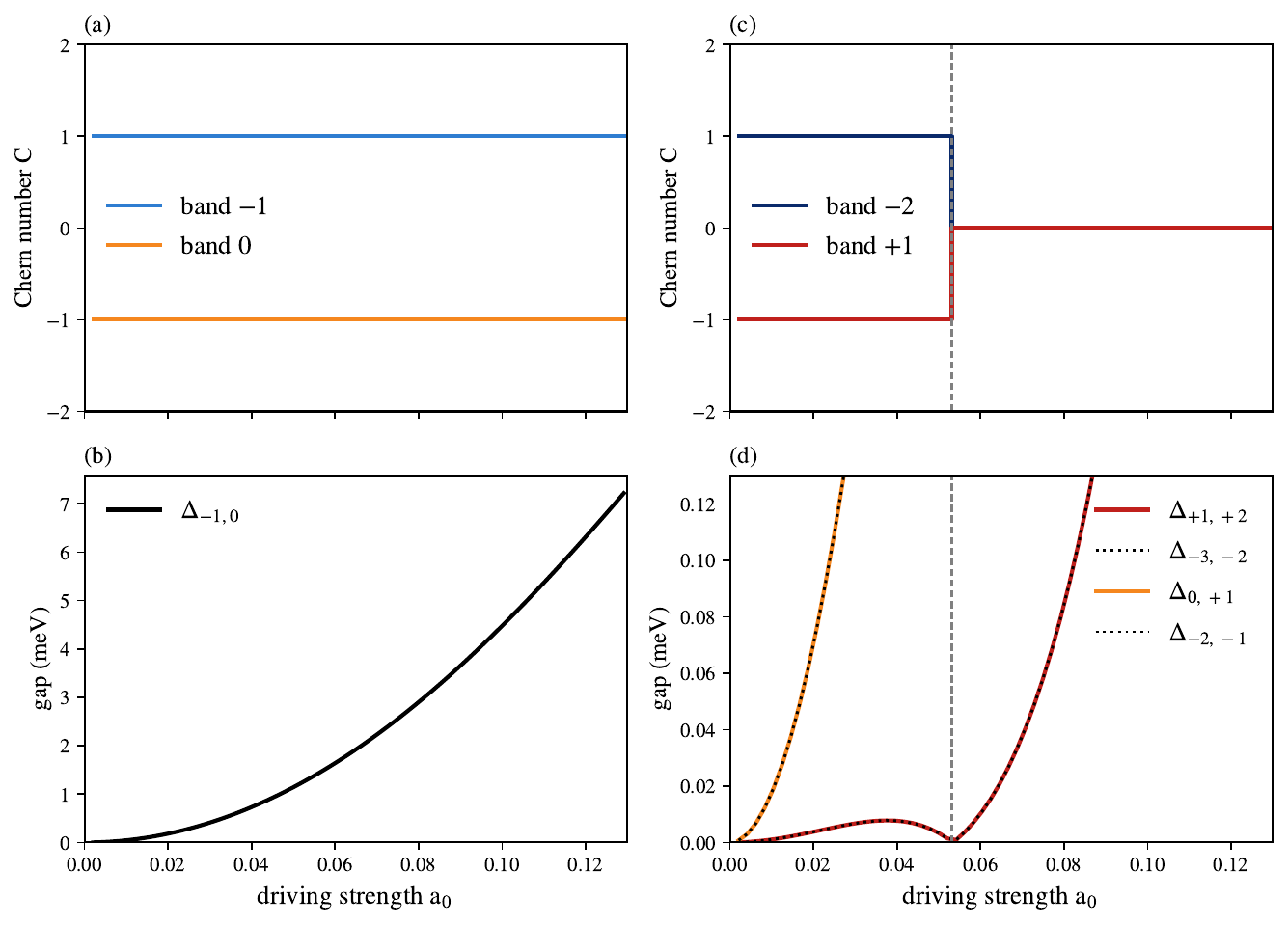}
    \caption{
  Line cut of the phase diagram of  Fig~\ref{phase_diagram_AA_CPL} at fixed period $L = 145$~nm.
  (a),(c)~Chern numbers of the inner bands $(-1,0)$ and the outer bands
  $(-2,+1)$ versus driving strength $a_0$. (b),(d)~Corresponding minimum direct
  gaps. The vertical dashed line marks the transition at which the outer bands
  become trivial, where the outermost gaps $\Delta_{-3,-2}$ and $\Delta_{+1,+2}$
  close.
}
    \label{aa_linecut}
\end{figure*}

We first discuss the topology of AB bilayer graphene under patterned
circularly polarized light. The phase diagram is shown in
Fig.~\ref{CPL_density}. The drive isolates the four central bands, labeled
band $(-2)$ through band $(+1)$ as defined in Fig.~\ref{Numbering}, so that
each carries a well-defined Chern number over almost the entire plane. As soon
as the drive is switched on, the two outer bands acquire Chern numbers $+2$ and
$-2$ while the two inner bands take $-1$ and $+1$, giving the pattern
$(+2,-1,+1,-2)$. This phase, in which every band is topological and the outer
bands carry the higher Chern number, occupies a broad region covering roughly
the lower-$a_0$ half of the plane. As the driving strength increases, the outer
bands each change by one unit, so that all four bands settle to
$(+1,-1,+1,-1)$, which fills the remaining, higher-$a_0$ region. The boundary
between the two phases shifts to smaller $a_0$ as the period $L$ grows, from
$a_0 \approx 0.09$ at $L = 100$~nm to $a_0 \approx 0.05$ at $L = 180$~nm,
because the modulation depth $M(L)$ increases with $L$ so that a weaker drive
suffices at larger period. Both phases occupy broad regions of the plane rather
than fine-tuned lines, so the sequence of transitions is robust.

At large driving strengths and long periods, all four bands show a narrow diagonal region where the Chern number changes and then returns to its previous value. This region approximately follows a line of constant $a_0L$. Starting from the unit-Chern phase $(+1,-1,+1,-1)$, the system first passes through a gap closing (as seen in Fig. \ref{linecute_AB}) into a narrow window where the Chern numbers reach $(-2,+2,-2,+2)$, and then, upon a second gap closing at slightly larger $a_0$, returns to $(+1,-1,+1,-1)$. Thus, this narrow region lies between two phase transitions. The phase with unit Chern numbers appears on both sides. The strip appears only for periods $L \gtrsim 150$~nm, extending approximately from $(a_0,L)=(0.15,150~\mathrm{nm})$ to $(0.12,180~\mathrm{nm})$. Its protecting gaps are only a few $\mu\mathrm{eV}$, indicating weak topological protection.

Overall, patterned circularly polarized light drives the AB bilayer from the $(+2,-1,+1,-2)$ phase at weak driving to the unit-Chern $(+1,-1,+1,-1)$ phase at stronger driving. Both phases occupy broad parameter ranges and should therefore be experimentally accessible. At the strongest driving and longest periods, a narrow re-entrant $(-2,+2,-2,+2)$ phase also appears, although observing it requires finer control of the driving parameters.

To connect the topological transitions to their gap closings, we take a line cut
at fixed period $L = 175$~nm, shown in Fig.~\ref{linecute_AB}. Panels (a) and (c)
show the Chern numbers of the inner and outer bands, and panels (b) and (d) the
corresponding direct gaps. Throughout the cut the inner gap $\Delta_{-1,0}$ in
panel (b) stays large, of order a few meV, so the two inner bands never touch
each other, every Chern change instead proceeds through a closing of one of the
inner--outer gaps, $\Delta_{-2,-1}$ or $\Delta_{0,+1}$, or of the outermost
gaps $\Delta_{-3,-2}$ or $\Delta_{+1,+2}$. At the main transition,
$a_0 \approx 0.05$, band $(-2)$ steps from $+2$ to $+1$ and band $(+1)$ from
$-2$ to $-1$ while the inner bands are unchanged; this coincides with a closing
of the outermost gaps $\Delta_{+1,+2}$ and $\Delta_{-3,-2}$ in panel (d), which
close together. At stronger driving, the system enters a narrow intermediate region at $a_0\approx0.122$ and leaves it at $a_0\approx0.130$. At each boundary, the Chern numbers of all four bands change simultaneously. On entering this region, the inner bands acquire Chern numbers $\pm2$, while the outer bands acquire $\mp2$; these changes are reversed on leaving. At both boundaries, the inner--outer gaps $\Delta_{-2,-1}$ and $\Delta_{0,+1}$ close. The inner gap $\Delta_{-1,0}$ remains open across the whole window. The inset in panel (d) resolves these two closings. Each Chern change in panels (a) and (c) is thus accompanied by a gap closing in panels (b) and (d), confirming the transition.

We now consider a different illumination scheme for AB bilayer graphene, combining a patterned longitudinal field of fixed strength $\eta_0=1.3$ with spatially uniform circularly polarized light of strength $a_0$. The resulting topological phase diagram is shown in Fig.~\ref{phase_AB_combined}. As soon as the circularly polarized component is switched on, it opens the protecting gaps, so the four central bands isolate immediately and acquire the Chern numbers $(+1,0,0,-1)$. This configuration holds over the entire $(a_0, L)$ plane studied; neither the outer nor the inner bands undergo any transition, and the phase is uniform throughout. Because there are no Chern changes anywhere in the diagram, a gap-closing line cut is not needed to understand the combined drive. The topology is fixed once the drive is switched on and does not evolve with either the driving strength or the period.

Thus, in AB-stacked bilayer graphene, the same four bands can realize different topological phases depending on the driving protocol. Under patterned circularly polarized light, the system passes through several broad topological phases. As the driving strength increases, the Chern numbers change from $(+2,-1,+1,-2)$ to the unit-Chern phase $(+1,-1,+1,-1)$. At the longest periods, a narrow intermediate region with Chern numbers $(-2,+2,-2,+2)$ appears between two unit-Chern regions. Each transition is accompanied by a gap closing. In contrast, the combined drive produces the phase $(+1,0,0,-1)$ throughout the parameter range shown. In this phase, only the outer bands are topological, while the inner bands remain trivial.

We next consider AA-stacked bilayer graphene under the influence of patterned circularly polarized light, using the same parameter space as in the AB-stacked case. The resulting topological phase diagram is shown in Fig.~\ref{phase_diagram_AA_CPL}. Here, the relevant bands, similar to the AB bilayer case, are the set of four bands surrounding the $+\gamma$ Dirac cone, labeled band $(-2)$ through band $(+1)$ as shown in Fig~\ref{Numbering}. As soon as the drive is switched on, these four bands isolate, and all are topological, with Chern numbers $(+1,+1,-1,-1)$. This phase occupies the low-$a_0$ region of the diagram. As the driving strength is increased, the two outer bands become trivial: band $(-2)$ drops from Chern number $+1$ to $0$ and band $(+1)$ from $-1$ to $0$, while the Chern numbers of the inner bands remain unchanged. The pattern is then $(0,+1,-1,0)$, which fills the remainder of the phase diagram. We emphasize that the inner bands keep Chern numbers $C = +1$ and $C = -1$ throughout, so the entire topological evolution is carried by the outer pair of bands, which passes from nontrivial to trivial at a single transition point. For $L\gtrsim150$~nm, the transition occurs near $a_0\approx0.055$ and depends only weakly on the period.

To connect the transition to its gap closing, we take a line cut at fixed period
$L = 145$~nm, shown in Fig.~\ref{aa_linecut}. Panels (a) and (c) give the Chern
numbers of the inner and outer bands, and panels (b) and (d) the corresponding
direct gaps. The inner gap $\Delta_{-1,0}$ in panel (b) opens smoothly and grows monotonically with $a_0$, and the inner bands hold $(+1,-1)$ across the whole cut. At $a_0\approx0.054$, the outermost gaps $\Delta_{-3,-2}$ and $\Delta_{+1,+2}$ close, as shown in panel (d). The outer bands therefore become trivial by touching neighboring bands outside the four-band manifold, rather than the inner bands. This gap closing coincides with the Chern-number change in panel (c), confirming the topological transition.

\begin{figure*}
    \centering
    \includegraphics[width=\linewidth]{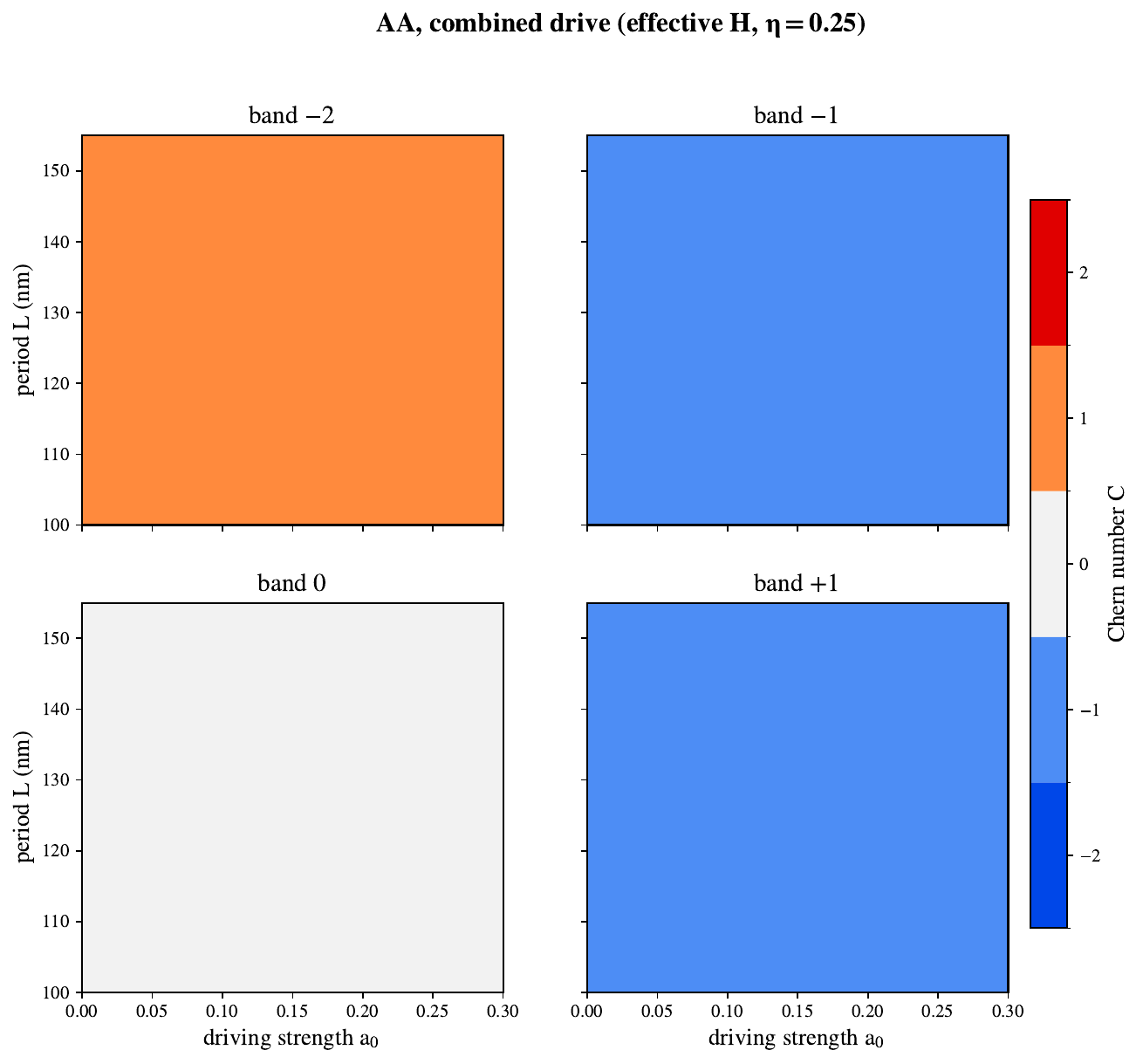}
    \caption{Phase diagrams of the four central bands of AA-stacked bilayer graphene
under the combined drive. The Chern numbers are shown as functions of the circularly
polarized driving strength $a_0 \in [0,0.3]$ and the superlattice period
$L \in [100,155]$~nm, with the longitudinal driving strength fixed at
$\eta_0 = 0.25$ and the driving frequency fixed at $\hbar\Omega = 3$~eV. The band
ordering, labeling, and color coding are the same as in Fig.~\ref{CPL_density}.}
    \label{phase_AA_combined}
\end{figure*}

Finally, we consider AA-stacked bilayer graphene under the influence of a combined drive consisting of a patterned longitudinal field with fixed strength $\eta_0=0.25$ and a uniform circularly polarized field of strength $a_0$. The resulting topological phase diagram is shown in Fig.~\ref{phase_AA_combined}. As soon as the drive is switched on, the four central bands surrounding the $+\gamma$ cone isolate and acquire the Chern numbers $(+1,-1,0,-1)$. This configuration holds over the entire parameter region of $(a_0, L)$ that was studied: none of the four bands undergoes a transition, and the phase is uniform throughout. Because no Chern number transitions occur anywhere in the diagram, a gap-closing line cut is not required for the combined drive; the topology is fixed once the drive is switched on and does not evolve with either the driving strength or the period. The four central Chern numbers sum to $-1$ rather than zero. We therefore also computed the Chern numbers of the two bands adjacent to the four-band manifold. The Chern numbers of these six central bands sum to zero, showing that changes in the four central bands are balanced by their nearest-neighbor bands.

Taken together, these results show that both the stacking and the driving protocol determine the topology of the four central Floquet bands. Patterned circularly polarized light produces several topological phases in AB stacking. In contrast, AA stacking after the initial gap opening undergoes a single transition in which only the outer bands change their Chern numbers. In contrast, the combined drive produces one robust phase over the full parameter range studied for each stacking. Notably, the Chern-number pattern still differs between AA and AB bilayers. The line-cut calculations, as a consistency check, further show that every Chern-number change is accompanied by a closing of the relevant band gap, as required.

\section{Conclusion}

In this work, we have shown that patterned electromagnetic fields can create a tunable optical superlattice in AA- and AB-stacked bilayer graphene, producing moir\'e-like minibands without a physical twist. We studied three driving protocols: patterned circularly polarized light, patterned longitudinal light, and a combined drive consisting of a patterned longitudinal field and a uniform circularly polarized field. The circularly polarized field couples to the intralayer Dirac motion, while the longitudinal field modulates the interlayer hopping.

We diagonalized the resulting space- and time-periodic Hamiltonians in a truncated Floquet--Bloch basis. For each driving protocol, we studied how the quasienergy spectrum evolves with driving strength, superlattice period, and driving frequency. Patterned circularly polarized light and the combined drive isolate the central bands in both stackings. In AB stacking, these bands also become nearly flat, whereas in AA stacking they remain dispersive even at strong driving. The longitudinal drive alone leaves the Dirac crossings intact and therefore does not isolate the central bands.

We used a Van Vleck high-frequency expansion to interpret these results. The circularly polarized field generates effective terms that open gaps at the Dirac points and modify the band dispersion. By contrast, the longitudinal drive preserves time-reversal symmetry and generates no corresponding effective mass term. Using the resulting effective Hamiltonian, we computed the Chern-number phase diagrams of the four central bands at fixed frequency. Under patterned circularly polarized light, the AB bands evolve from $(+2,-1,+1,-2)$ at weak driving to $(+1,-1,+1,-1)$ at stronger driving, with an additional narrow intermediate phase at the longest periods. In AA stacking, only the outer bands undergo a transition and become topologically trivial. Under the combined drive, each stacking exhibits a single topological phase throughout the parameter range studied.

Patterned electromagnetic fields therefore provide a tunable platform for studying moir\'e-like physics without a fixed twist angle. The superlattice period and strength can be controlled through the mask geometry and field intensity, respectively, allowing control over band isolation, dispersion, and topology. The same approach may also apply to other layered materials.

A natural extension is to study the light-induced superlattice in a perpendicular magnetic field. The Hofstadter regime is reached when the magnetic flux through a superlattice unit cell becomes comparable to the flux quantum. At experimentally accessible magnetic fields, this condition favors large superlattice periods. Because the optical period can be selected through the mask geometry, patterned driving may provide a tunable route to Hofstadter physics in bilayer graphene.
 \section{Acknowledgments}
 M.V. gratefully acknowledges the support provided by the Deanship of Research Oversight and Coordination (DROC) and the Interdisciplinery Reasearch Center(IRC) for Advanced Quantum Computing (AQC) at King Fahd University of Petroleum \& Minerals (KFUPM) for funding their contribution to this work through research grant No. INQC2607.

\bibliographystyle{unsrt}
\bibliography{literature}

\appendix
\section{Frequency and length dependence}
\begin{figure*}
    \centering
    \includegraphics[width=1.0\linewidth]{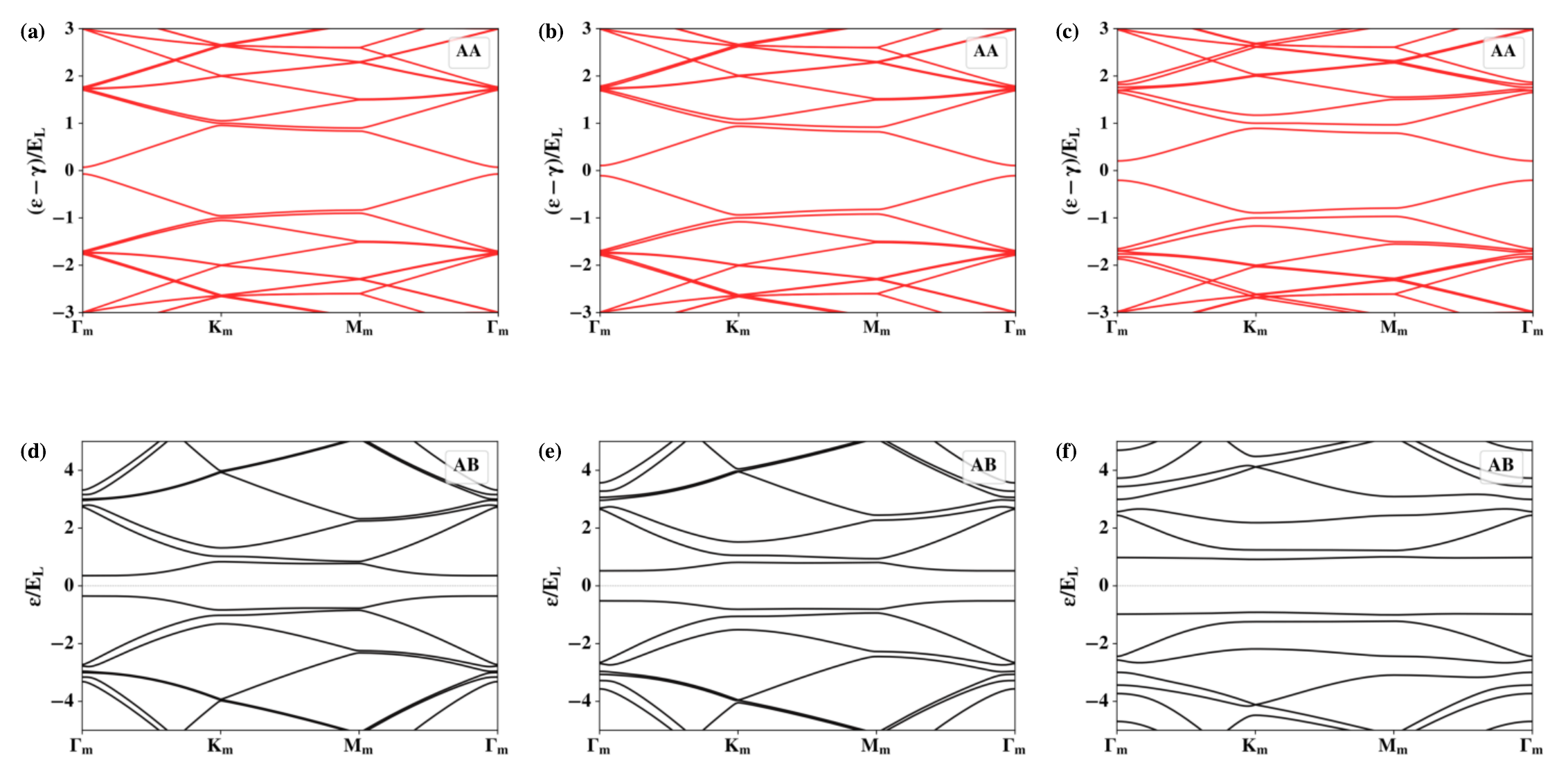}
    \caption{Quasienergy band structures under the patterned circularly polarized drive
at decreasing drive frequency. Panels (a)-(c) show AA-stacked bilayer graphene
with $a_0=0.1$ for $\hbar\Omega=3$, $2$, and $1.0$~eV, respectively; panels
(d)-(f) show AB-stacked bilayer graphene with $a_0=0.06$ for the same frequencies.
In all cases $L=100$~nm. The quasienergy is plotted in units of $E_L$, defined
separately for the two stackings as described in the text.} 
    \label{fig:CPL_F}
\end{figure*}
\begin{figure*}
    \centering
    \includegraphics[width=\linewidth]{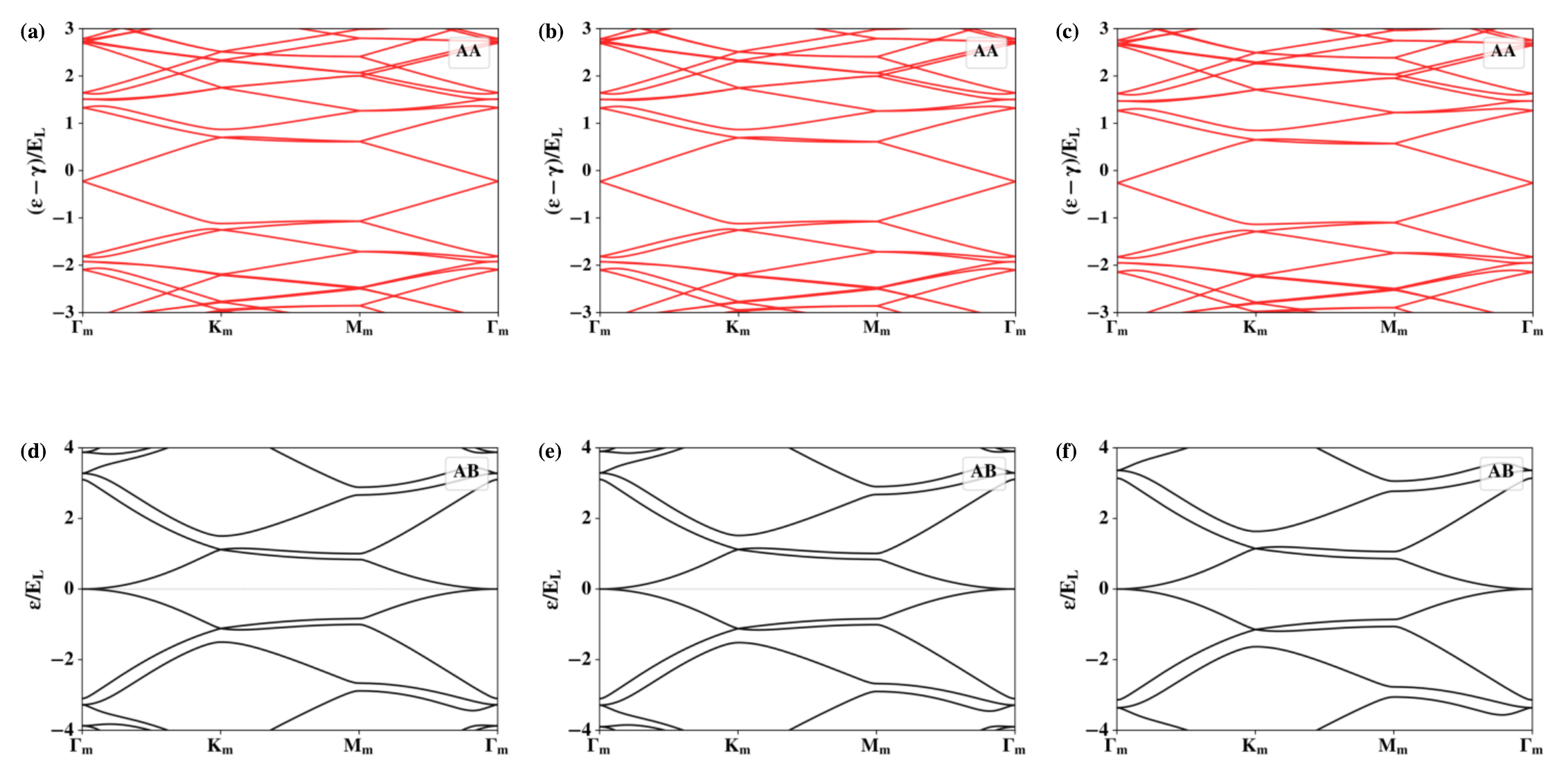}
    \caption{Quasienergy band structures under the patterned longitudinal drive
at decreasing drive frequency. Panels (a)-(c) show AA-stacked bilayer graphene
with $\eta_0=0.3$ for $\hbar\Omega=3$, $2$, and $1.0$~eV, respectively; panels
(d)-(f) show AB-stacked bilayer graphene with $\eta_0=0.8$ for the same frequencies.
In all cases $L=100$~nm. The quasienergy is plotted in units of $E_L$, defined
separately for the two stackings as described in the text. For
AA and AB, the truncation was enlarged to maintain convergence
 ($N_s=5$, $N_F=5$ for AA),$N_s=3$, $N_F=5$ for AB)}
    \label{LOG_F}
\end{figure*}

\begin{figure*}
    \centering
    \includegraphics[width=\linewidth]{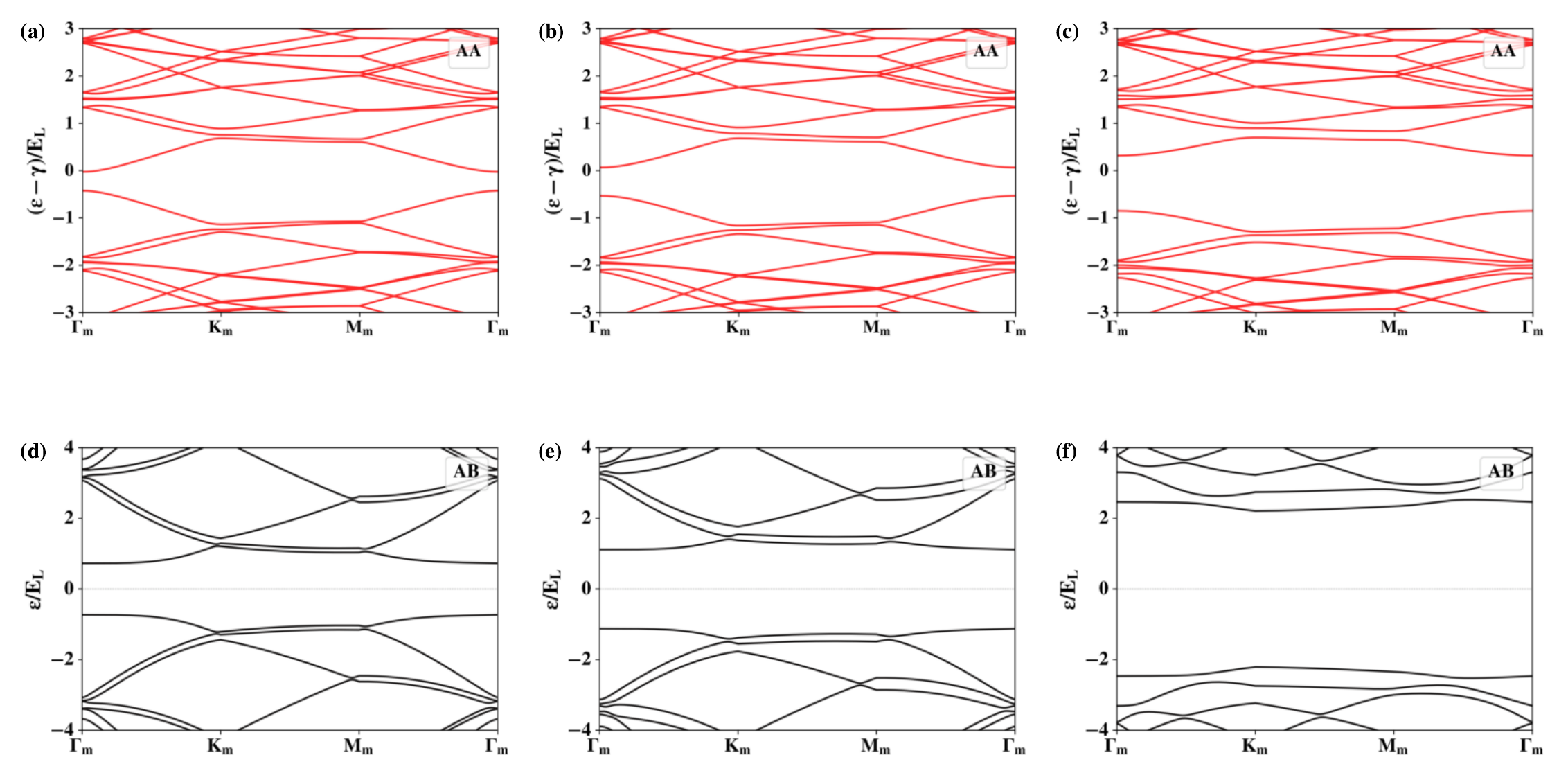}
    \caption{Quasienergy band structures under the combined drive: patterned longitudinal drive plus uniform circularly polarized light 
at decreasing drive frequency. Panels (a)-(c) show AA-stacked bilayer graphene
with fixed $\eta_0=0.3$, $a_0=0.2$ for $\hbar\Omega=3$, $2$, and $1.0$~eV, respectively; panels
(d)--(f) show AB-stacked bilayer graphene with fixed $\eta_0=0.5$, $a_0$=0.10 for the same frequencies.
In all cases $L=100$~nm. The quasienergy is plotted in units of $E_L$, defined
separately for the two stackings as described in the text. For
AA, the truncation was enlarged to maintain convergence
 ($N_s=5$, $N_F=5$ for AA).}
    \label{Log_cpl_f}
\end{figure*}

\begin{figure*}
    \centering
    \includegraphics[width=\linewidth]{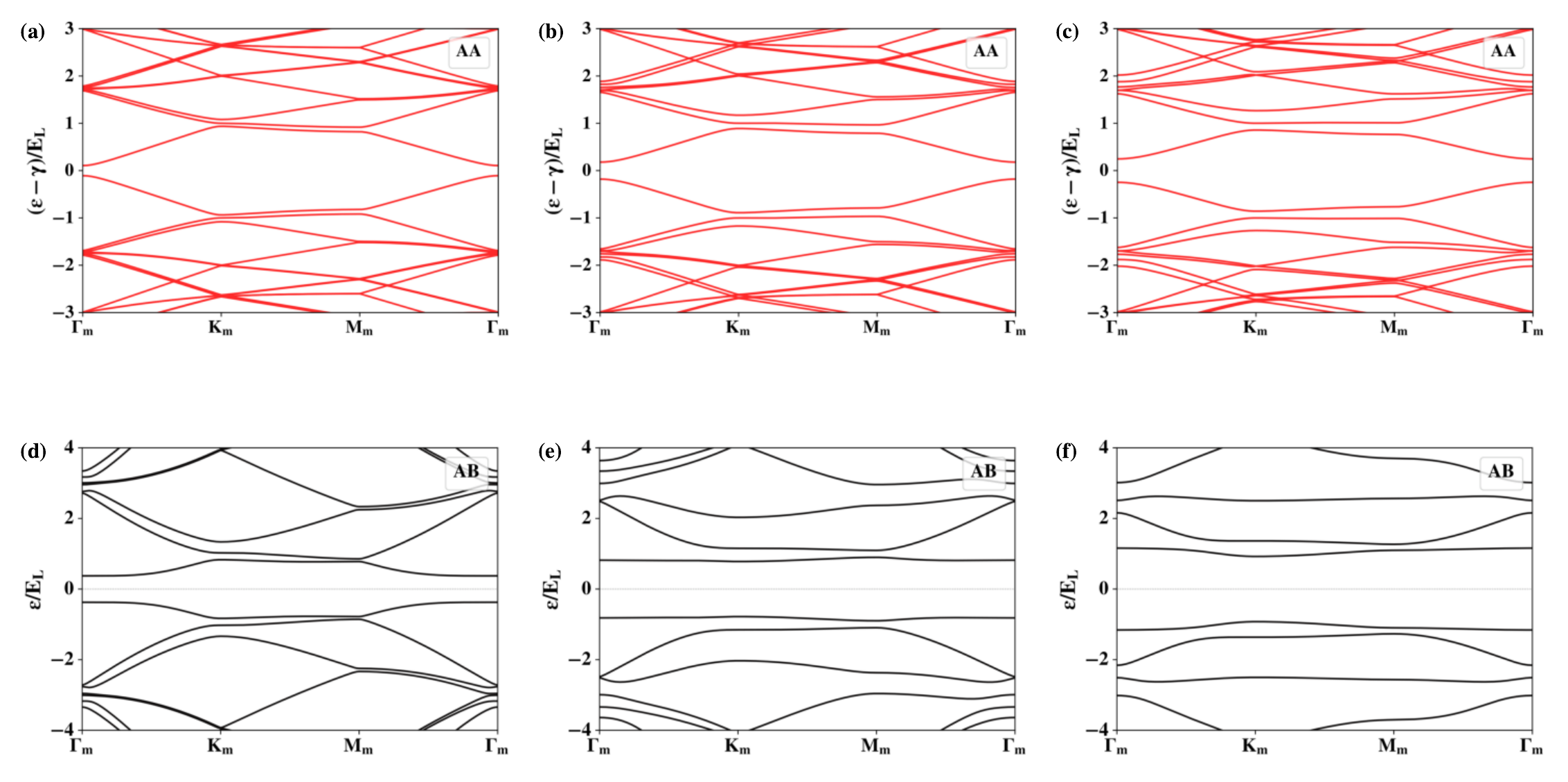}
    \caption{Quasienergy band structures under the patterned circularly polarized
drive at fixed drive frequency and increasing superlattice period. Panels (a)-(c)
show AA-stacked bilayer graphene at fixed $a_0=0.1$ and panels (d)-(f) show AB-stacked bilayer graphene at fixed $a_0=0.05$, for $L=100$, $150$, and $200$ nm. In all cases, $\hbar\Omega=2$ eV. The quasienergy is plotted in units of $E_L$, defined separately for the two stackings as described in the text.}
    \label{CPL_L}
\end{figure*}

\begin{figure*}
    \centering
    \includegraphics[width=\linewidth]{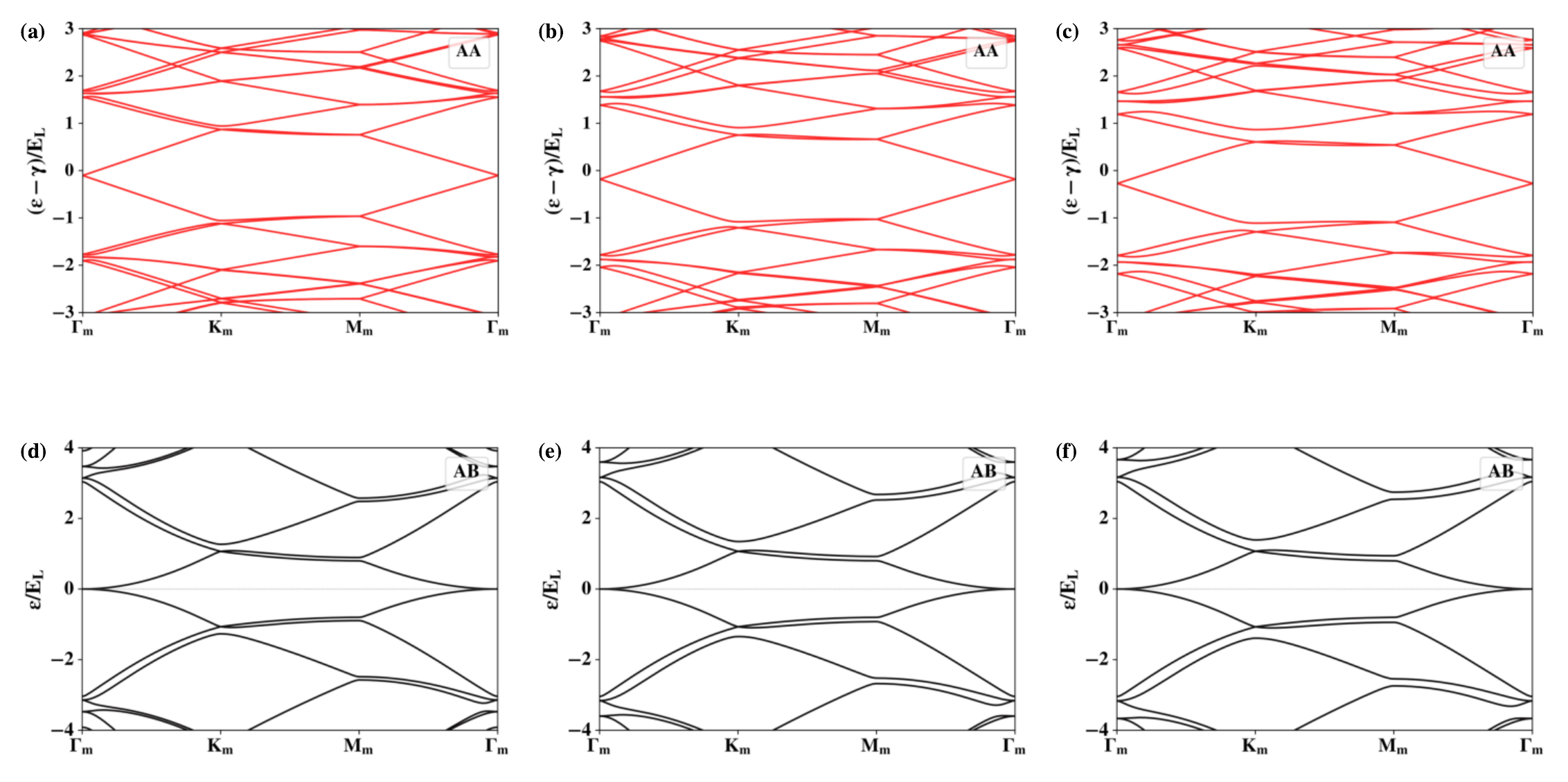}
    \caption{Quasienergy band structures under the patterned longitudinal drive at fixed frequency and increasing superlattice period. Panels (a)-(c)
show AA-stacked bilayer graphene at fixed $\eta_0=0.2$ and panels (d)-(f) show AB-stacked bilayer graphene at fixed $\eta_0=0.6$, for $L=100$, $150$, and $200$ nm. In all cases, $\hbar\Omega=2$ eV. The quasienergy is plotted in units of $E_L$, defined separately for the two stackings as described in the text. For AA the truncation was enlarged to maintain convergence($N_s=5$, $N_F=5$ for AA).}
    \label{LOG_L}
\end{figure*}

\begin{figure*}
    \centering
    \includegraphics[width=\linewidth]{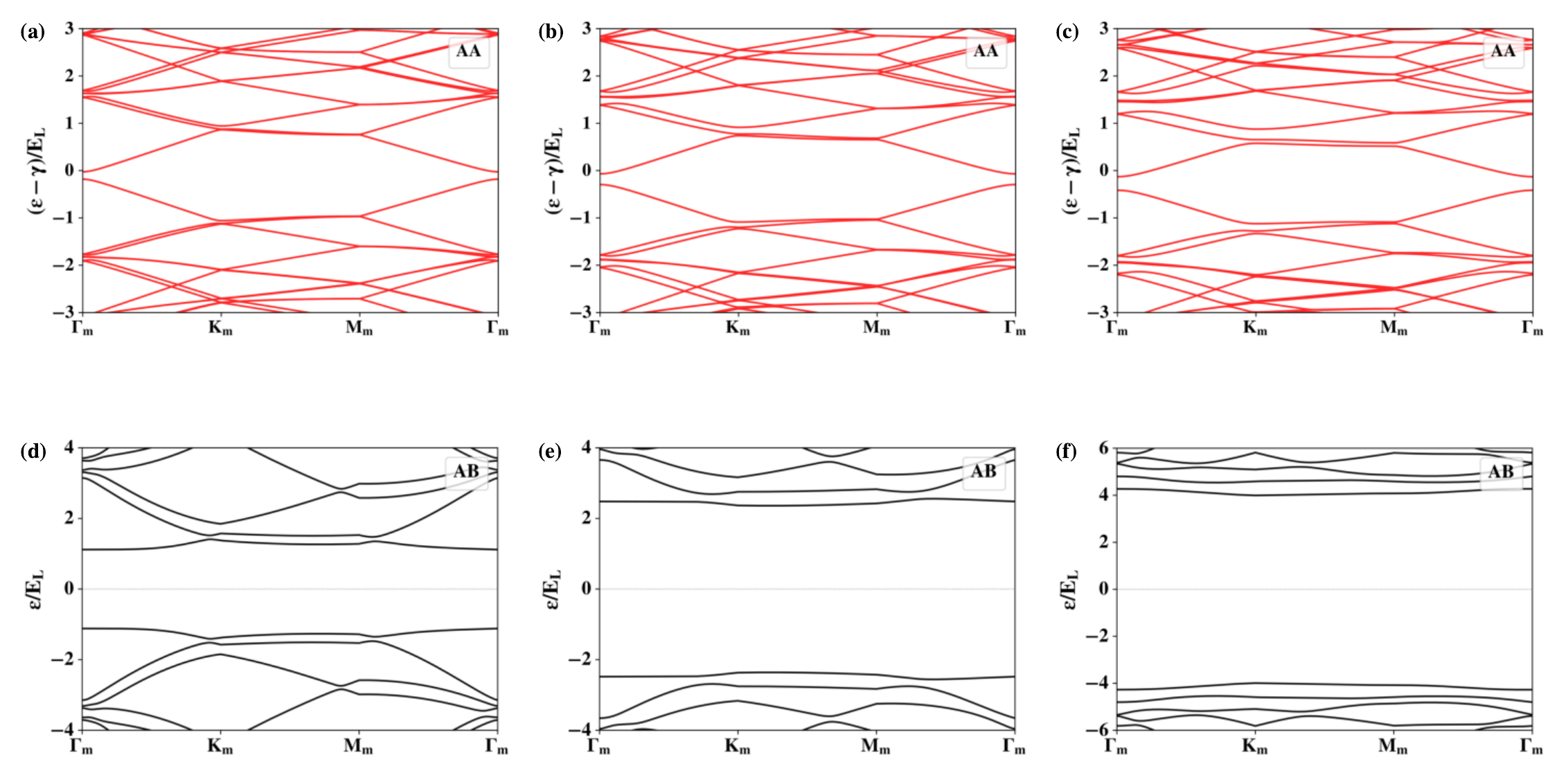}
    \caption{Quasienergy band structures under the combined drive of patterned longitudinal light and uniform circularly polarized light at fixed frequency and increasing superlattice period. Panels (a)-(c) show AA-stacked bilayer graphene at fixed $\eta_0=0.1$ and $a_0=0.1$, and panels (d)-(f) show AB-stacked bilayer graphene at fixed $\eta_0=0.6$ and $a_0=0.1$, for $L=100$, $150$, and $200$ nm. In all cases, $\hbar\Omega=2$ eV. The quasienergy is plotted in units of $E_L$, defined separately for the two stackings as described in the text. For AA the truncation was enlarged to maintain convergence($N_s=5$, $N_F=5$ for AA).}
    \label{LOG_CPL_L}
\end{figure*}
In the main text, we examined the effect of varying the driving strength while keeping the driving frequency and superlattice period fixed. Here we present the complementary parameter sweeps, in which the driving frequency is reduced at fixed period and the superlattice period is increased at fixed frequency. We place these results in the Appendix because they exhibit the same qualitative behavior as increasing the driving strength, as discussed in the results, and we include them primarily to provide the complete parameter dependence. In all calculations, the Floquet-Bloch Hamiltonian satisfies the convergence criterion described in the main text, and the driving frequency remains larger than the spectral width of the truncated Hamiltonian, ensuring that the Floquet replicas do not overlap the central bands.
It is interesting to note that the frequency and length sweep, from 3 to 2 to 1~eV, from 100~nm to 200~nm, reproduces the same qualitative behavior observed when the driving strength is increased, as shown in Figs.~\ref{CPL_L}, \ref{LOG_L}, and \ref{LOG_CPL_L}. As the frequency is reduced, the central bands become progressively flatter and more isolated. By $\hbar\omega=1$~eV, the AB bilayer develops an isolated pair of nearly flat bands under both the patterned circularly polarized drive and the combined drive, with the latter providing the closest realization of a driven moir\'e  material. In contrast, the longitudinal drive alone leaves the Dirac crossings protected throughout the entire frequency range. From an experimental perspective, this provides an alternative route to the flat-band regime, since comparable band structures can be achieved by lowering the driving frequency rather than increasing the laser intensity, thereby reducing the required optical power.

\end{document}